\documentclass{ws-ijmpd}
\usepackage[super,compress]{cite}
\usepackage{graphics} 
\usepackage{amssymb}
\usepackage{bm}
\usepackage{wasysym}

\begin{document}

\markboth{Rainer Dick}
{Direct signals from electroweak singlets through the Higgs portal}

\title{DIRECT SIGNALS FROM ELECTROWEAK SINGLETS THROUGH THE HIGGS PORTAL
}

\author{RAINER DICK
}

\address{Department of Physics and Engineering Physics, 
University of Saskatchewan,\\
116 Science Place,
Saskatoon, SK S7N 5E2, Canada
\\
rainer.dick@usask.ca}

\maketitle

\begin{abstract} 
We review predictions and constraints for nuclear recoil 
signals from Higgs portal dark matter under the assumption of
standard thermal creation from freeze-out.
Thermally created scalar and vector Higgs portal dark matter masses 
are constrained to be in the resonance region near half the
Higgs mass, $m\lesssim m_h/2$, or above several TeV. The resonance region
for these models will be tested by XENONnT and LZ. The full mass range
up to the unitarity limit can be tested by DarkSide-20k and DARWIN.
 Fermionic Higgs portal dark matter with a pure CP odd coupling is constrained 
by the Higgs decay width, but has strongly 
suppressed recoil cross sections which cannot be tested with upcoming experiments.
 Fermionic Higgs portal dark matter with a combination of CP even and odd
Higgs couplings can be constrained by the direct search experiments.
\end{abstract}


\ccode{PACS numbers: 95.35.+d; 12.60.Fr; 14.80.Bn}

\section{Introduction}	
\label{sec:intro}

The existence of a dark sector is firmly established through many 
different astronomical techniques. 
It is now widely accepted that cold dark 
matter forms and sustains large scale structure in the universe because 
the amount of baryonic matter is severely constrained by the successful theory 
of big bang nucleosynthesis and by microlensing searches.
The existence of dark matter is also
directly inferred from galactic rotation curves, motion of galaxies in 
galaxy clusters, gravitational lensing, gas temperature in galaxy clusters,
and from the observational separation of subdominant baryonic mass components
and dominant dark halos in cluster collisions\cite{clusters1,clusters2}. For 
indirect evidence for dark matter, we know
that baryons alone could not have generated the observed large scale structure
in the universe without the dominant gravitational pull and the head start from
cold dark matter, which could start to clump together under the influence of
gravity well before the primordial plasma recombined to form neutral atoms.
The co-existence of the uniformly distributed dark energy component and
the cold dark matter halos around galaxies and galaxy clusters is also confirmed 
through type 1a supernova fits to distance-redshift relations\cite{SN1a1,SN1a2} 
and through fits of cosmological parameters to temperature fluctuations in the 
cosmic microwave background\cite{planck}, see also Ref.~\refcite{freese} for
a recent overview.

 Furthermore, if we make the reasonable assumption that the cold dark matter 
abundance was generated through the same thermal production mechanisms in the early
universe which determined baryonic abundances (after the emergence of 
matter--anti-matter asymmetry in the baryonic sector) from one common heat bath, 
we can infer that dark matter and baryonic matter should not only interact 
gravitationally, but also e.g. through weak particle interactions. It then appears
likely that we should also directly detect dark matter particles in particle
physics laboratories, either through production from standard matter collisions at 
the Large Hadron Collider or future accelerators, or through nuclear recoil searches
at underground labs like SNOLab, the Laboratori Nazionali del Gran Sasso, the Sanford 
Underground Research Facility, or the China JinPing Underground Laboratory.

Dark matter research therefore holds the promise to provide us with a window into
particle physics beyond the Standard Model. From a theoretical perspective this begs
the question: What do we expect from particle physics beyond the Standard Model?
We can try to connect dark matter theory with theoretical attempts to solve other
contemporary problems of fundamental physics, e.g. the problem of quantum gravity
or the problem to explain the huge difference between the electroweak scale and
the Planck scale. We can therefore try to identify candidates for dark matter 
in string theory, e.g. in the gravitational sector or the hidden $E_8$ sector
of heterotic string theory, or we can connect it to enhanced symmetries in
particle physics through supersymmetric models or Grand Unified Theories
or left-right symmetric models.
A different approach tries to focus on the problem at hand and only expand
the Standard Model with a minimal number of additional helicity states to
explain dark matter. These models are known as \textit{minimal dark matter
models}. However, extending the Standard Model with a small number of
helicity states to account for dark matter generically will not imply stability
of this extended Standard Model up to the Planck scale, although Higgs portal 
couplings can improve the stability properties of the Standard 
Model\cite{gonderinger1,profumo,gonderinger2,lebedev1,petcov,khoze,khan,lebedev2,garg}.
Nevertheless, an appealing property of minimal models (besides Occam's razor) is 
their predictive power and ensuing testability in particle physics experiments.
Indeed, we will see that results from ATLAS, CMS, XENON1T,
LUX, and PandaX-II put already very strong constraints on several minimal Higgs
portal models if we assume standard thermal freeze-out. Besides being testable
in direct and collider based search experiments,
the predictive power of these models is also appealing from the discovery 
perspective: The direct relation between dark matter mass and Higgs portal
coupling also implies that any potential signal from direct search experiments
determines both the coupling and the mass of the dark matter particles.
The signal can therefore not only be tested by other direct search experiments
(and in the low mass sector also by ATLAS and CMS), but also by gamma-ray 
and neutrino telescopes by focusing on the energy range of annihilation 
products corresponding to the proposed dark matter mass.

It was in the framework of the minimal models that attention turned in particular to 
virtual Higgs exchange as a promising coupling between dark matter and baryons and 
where the phrase \textit{Higgs portal} for such a coupling was coined \cite{frank}. 
However, Higgs exchange is not limited to minimal dark matter models, but can also 
occur in many of the more elaborate models.

Indeed, it is very natural to expect virtual Higgs exchange to contribute to
interactions between baryons and dark matter. With mounting evidence that 
neutrinos also have mass, it appears increasingly likely that the Higgs particle
couples already at tree-level to every Standard Model particle except photons
and gluons. Therefore, besides gravity, Higgs exchange is the only known interaction 
which affects almost every known particle directly through Yukawa couplings. 
While neither necessary nor unavoidable, the assumption that 
dark matter particles also couple to the Higgs field certainly appears natural,
and it is these kinds of models and their prospects for direct search experiments,
which are the focus of this review.

Over the years
numerous thermal and non-thermal mechanisms have been proposed for the 
generation of dark matter in the early universe. Creation through a Higgs portal 
can e.g. play a role in the freeze-in of dark matter components which are 
interacting so weakly that they are never 
thermalized\cite{fi1,fi2,fi3,fi4,fi5,fi6,fi7,fi8,fi9,fi10,fi11,fi12,fi13}.
 Furthermore, Higgs portal interactions can contribute to the interactions
of asymmetric dark 
matter\cite{asym1,asym2,asym3,asym4,asym5,asym6,asym7,asym8,asym9,asym9b,asym10,asym11,asym12}.
Thermal freeze-out from a heat bath is a generic 
mechanism for creating relic matter abundances\cite{LW,KT,dodelson}, and the 
question arises whether it is the 
dominant or a sub-dominant mechanism for the creation of dark matter in the early 
universe. Thermal freeze-out would e.g. be negligible if the coupling of the
potential dark matter component is too large, which would result in too large
of a thermally averaged annihilation cross section\cite{gondolo} 
\begin{equation}\label{eq:thermalGG}
\langle\sigma v\rangle(g,T)=\frac{1}{8m^4 TK_2^2(m/T)}
\int_{4m^2}^\infty\!ds\,\sqrt{s}\left(s-4m^2\right)\sigma(s)K_1(\sqrt{s}/T)
\end{equation}
of the component. Here $s$-channel domination is assumed and the dependence on 
dark matter couplings $g$ is implicit in $\sigma(s)$.
Large annihilation cross section means late decoupling
and low remnant freeze-out density $\varrho\propto\langle\sigma v\rangle^{-1}$ 
of the component. In that case the component has to be generated by other means,
e.g. through phase transitions or coherent oscillations, to become 
a viable dark matter candidate. 

Mass constraints from direct detection limits can be relaxed in models with
effective Sommerfeld enhancement of the annihilation cross sections\cite{se1,se2,se3},
or through co-annihilation terms which
prevent early freeze-out for small dark matter coupling\cite{coann1,coann2,casas}. 
However, this is not the case in the minimal $\mathbb{Z}_2$ symmetric Higgs portal 
models. 
Here we will define a minimal Higgs portal model as a model where 
standard thermal freeze-out is the relevant mechanism for dark matter creation, 
and all assertions about experimentally ruled out or permitted dark matter mass
values are only valid under that premise. 

Thermal freeze-out models for dark matter are constrained by unitarity\cite{GK} 
to dark matter masses $m_D\lesssim 100$ TeV. This arises from the fact that
unitarity of the scattering matrix in partial wave expansion implies a bound
on the total scattering cross section\cite{weinberg},
\begin{equation}\label{eq:unit1}
\sigma\le\frac{4\pi}{k^2(2s_1+1)(2s_2+1)}
\sum_{\ell=0}^{\infty}\sum_{s=|s_1-s_2|}^{s_1+s_2}\sum_{j=|\ell-s|}^{\ell+s}
(2j+1),
\end{equation}
where $s_1$ and $s_2$ are the spins of the two incoming particles. In particular,
the bound on $s$-wave dominated annihilation cross sections 
is $\sigma^{(\ell=0)}\le 4\pi/k^2$, and this yields in the non-relativistic
limit
\begin{equation}\label{eq:unit2}
v\sigma^{(\ell=0)}\le\frac{4\pi}{m_D^2 v}. 
\end{equation}
This upper limit implies for the low-energy dominated thermally averaged
annihilation cross section of the dark matter particles
the estimate $\langle v\sigma\rangle\lesssim (4\pi/m_D^2)\langle v^{-1}\rangle$.
On the other hand, the required thermally averaged annihilation cross section for
generating the observed cold dark matter abundance from freeze-out 
$\langle v\sigma\rangle\lesssim 3\times 10^{-26}\,\mathrm{cm}^3/\mathrm{s}$ 
varies very weakly with mass. Estimates for the 
parameter $\langle v^{-1}\rangle$ then yield
the upper limit\cite{GK} $m_D\lesssim 100$ TeV.

 Furthermore, the requirement of
perturbativity can imply additional constraints on the high mass values, since 
the required thermal cross section $\langle\sigma v\rangle_f$ for dark 
matter freeze-out varies only logarithmically with mass, whereas the actual
thermally averaged annihilation cross section $\langle\sigma v\rangle(g,T_f)$
is suppressed for large dark matter masses. This implies that the required
dark matter 
couplings $g^2=\langle\sigma v\rangle_f/\langle\sigma v\rangle(1,T_f)$
increase with dark matter mass, and for the bosonic models they eventually reach 
the (conventionally defined) non-perturbative limit $g\simeq 4\pi$ before the 
unitarity limit is reached. This happens at masses of several ten TeV, 
and means that perturbative calculations of recoil cross sections cannot be 
trusted above the perturbativity mass limits and should only be considered as 
order of magnitude estimates near the unitarity limit for the bosonic models. 
The estimates are nevertheless interesting guide posts for the power of current 
and future direct search experiments to push particle physics to the unitarity 
limit, and therefore we will also display those leading order estimates for the 
bosonic recoil cross sections. 

The sensitivities of the Argon and Xenon based direct search experiments are 
extrapolated into the high mass regime, usually up to 10 TeV for XENON1T, and up to
the unitarity limit $m_D\lesssim 100$ TeV for future experiments. Far above the 
detector threshold, the number of recoil events scales with dark matter mass
like $m_D^{-1}S(q(m_D))$ such that exclusion limits on the basis of the same
number of expected events scale like $m_DS^{-1}(q(m_D))$. The first factor is due
to the reduced dark particle flux $j=\varrho_Dv_D/m_D$. The second factor is due  to 
the decrease of the nuclear structure factor\cite{lewin,vietze} $S(q)=F^2(q)$ with 
momentum, which is taken into account through a Helm structure factor. The maximal 
nuclear recoil energy and maximal momentum transfer in Xenon 
are $E_r\lesssim 175$ keV and $Q\lesssim 200\,\mathrm{MeV}/c$, respectively. 
The minimal exchanged Higgs wavelength is $\lambda\gtrsim 6$ fm, such that the 
virtual Higgs quanta still probe the nuclei.
Actual mass reconstruction for very heavy
dark matter from a recoil signal will be challenging\cite{darwin} because the 
momentum transfer $q$ depends on $m_D$ only through the reduced mass of the 
scattering partners, but this does not prevent direct search experiments from 
pushing the limits far beyond 10 TeV, and indeed all the way up to
the Planck scale\cite{bblr}.

The theory of the effective Higgs nucleon coupling and the impact of strangeness
in the nucleon will be reviewed in Sec.~\ref{sec:svz}. Scalar, vector, 
and fermionic Higgs portal models will then be reviewed in 
Secs.~\ref{sec:scalar}, \ref{sec:vector} and \ref{sec:fermions}, respectively.

\section{The Higgs-nucleon coupling}	
\label{sec:svz}

Higgs portals entail that dark matter interacts with ordinary matter through Higgs exchange.
Shifman, Vainshtein and Zakharov\cite{SVZ} had demonstrated that the effective Higgs-nucleon 
coupling $g_{hN}h\overline{N}N$ has at least a strength
\begin{equation}\label{eq:gvhsvz}
g_{hN}v_h\simeq 210\,\mathrm{MeV},
\end{equation}
where $v_h=246$ GeV is the vacuum expectation value of the Higgs field. The Higgs-nucleon 
coupling is therefore much larger than the Higgs-electron
coupling $g_{he}v_h=511$ keV. Since recoil cross sections for dark matter mass $m_D$, 
Higgs-to-dark matter
coupling $g_{hD}$, and Higgs-to-standard particle $P$ coupling $g_{hP}$ are of order
$\sigma_{\mathrm{rec}}\simeq g^2_{hD}g^2_{hP}v_h^2m_P^2/\pi m_h^4(m_D+m_P)^2$, only nucleon
recoils are relevant for Higgs portal dark matter.

However, the Higgs-nucleon coupling depends critically on the strangeness content
of the nucleon, which is often expressed in terms of the $y$-parameter
\begin{equation}\label{eq:yN}
y_N=\frac{2\langle N|\overline{s}s|N\rangle}{
\langle N|\overline{u}u+\overline{d}d|N\rangle}
=\frac{m_u+m_d}{m_s}\frac{\sigma_{sN}}{\sigma_{\pi N}}
\simeq\frac{m_u+m_d}{m_s}\frac{f_s^N}{f_u^N+f_d^N},
\end{equation}
where
\begin{equation}\label{eq:sigmasN}
\sigma_{sN}=m_s\langle N|\overline{s}s|N\rangle=m_N f_s^N
\end{equation}
and
\begin{equation}\label{eq:sigmapiN}
\sigma_{\pi N}=\frac{m_u+m_d}{2}\langle N|\overline{u}u+\overline{d}d|N\rangle
\simeq m_N(f_u^N+f_d^N)
\end{equation}
are known as $\sigma$-parameters of the nucleon and $m_N$ is the nucleon mass. 
The parameters $f_q^N$ are defined as $f_q^N=m_q\langle N|\overline{q}q|N\rangle/m_N$.

Shifman, Vainshtein and Zakharov
had used the fact that the $\sigma$-parameter $\sigma_{\pi N}$ is much smaller
than the nucleon mass $m_N$ and worked under the assumption of negligible 
strangeness in the nucleon, $y_N\ll 1$. In this case the 
effective Higgs-nucleon coupling is dominated by the coupling of the Higgs
to the three heavy quark species charm, bottom and top,
\begin{equation}\label{eq:gvh1}
g_{hN}v_h\Big|_{y_N\ll 1}\simeq\sum_{Q=c,b,t}m_Q\langle N|\overline{Q}Q|N\rangle,
\end{equation}
where the heavy condensates arise from virtual fluctuations of
the gluon sea in the nucleon. However, the strangeness content of the nucleon 
is still not very well known and can have an appreciable effect on the 
effective Higgs-nucleon coupling. This directly affects the predictions for 
direct dark matter signals\cite{ellis,goudelis,rdfs,hoferichter}, as 
explained in Eq.~(\ref{eq:gvh4}) below. 

Cheng\cite{tpcheng} e.g. had argued for a large scalar 
strange form factor $y_N\simeq 0.47$. This would increase the effective
Higgs-nucleon coupling to $g_{hN}v_h=530$ MeV because the relatively strong
coupling of the Higgs to the strange quark overcompensates for the 
reduction of the heavy quark condensates which in turn arises from a 
reduction of the gluon content. Cheng's estimate would increase nuclear recoil 
cross sections of Higgs portal dark matter by more than a factor of 6, and would 
rule out much larger mass ranges in particular for the bosonic Higgs portal models. 
However, there is wide consensus now that $y_N$ is small.
We will follow and update the recent analysis performed in Ref.~\refcite{rdfs}, 
and find a (very cautious) current uncertainty of order 2 in Higgs portal 
recoil cross sections.

The coupling of the Higgs field to the quarks in the nucleon yields an effective 
Higgs-nucleon coupling in the form
\begin{equation}\label{eq:gvh2}
g_{hN}v_h=\sum_{q=u,d,s}m_q\langle N|\overline{q}q|N\rangle
+\sum_{Q=c,b,t}m_Q\langle N|\overline{Q}Q|N\rangle.
\end{equation}
On the other hand, the trace anomaly leads to an equation for the nucleon mass 
in terms of quark and gluon content\cite{SVZ},
\begin{equation}\label{eq:MN1}
m_N=\sum_{q=u,d,s}m_q\langle N|\overline{q}q|N\rangle
+\sum_{Q=c,b,t}m_Q\langle N|\overline{Q}Q|N\rangle
-\frac{7\alpha_s}{8\pi}\langle N|G^2|N\rangle,
\end{equation}
while effective heavy quark theory yields for the heavy quark species
\begin{equation}\label{eq:mQ}
m_Q\langle N|\overline{Q}Q|N\rangle=-\frac{\alpha_s}{12\pi}
\langle N|G^2|N\rangle.
\end{equation}
We can combine Eqs. (\ref{eq:yN},\ref{eq:sigmapiN},\ref{eq:gvh2},\ref{eq:mQ}) to
eliminate the scalar quark form factors from $m_N$ and $g_{hN}v_h$,
\begin{equation}\label{eq:MN2}
m_N\simeq\sigma_{\pi N}+\frac{m_s y_N}{m_u+m_d}\sigma_{\pi N}
-\frac{9\alpha_s}{8\pi}\langle N|G^2|N\rangle
\end{equation}
and
\begin{equation}\label{eq:gvh3}
g_{hN}v_h\simeq\sigma_{\pi N}+\frac{m_s y_N}{m_u+m_d}\sigma_{\pi N}
-\frac{\alpha_s}{4\pi}\langle N|G^2|N\rangle.
\end{equation}
This yields an expression for the effective Higgs-nucleon coupling $g_{hN}v_h$ 
in terms of $\sigma_{\pi N}$, $y_N$, and masses \cite{rdfs},
\begin{eqnarray}\nonumber
g_{hN}v_h&\simeq&\frac{7}{9}\left(1+\frac{m_s y_N}{m_u+m_d}\right)\sigma_{\pi N}
+\frac{2}{9}m_N
\\ \label{eq:gvh4}
&\simeq&\frac{7}{9}\sum_{q=u,d,s}m_N f_q^N+\frac{2}{9}m_N
\equiv m_N f_N.
\end{eqnarray}
The assumptions of Shifman, Vainshtein and Zakharov imply negligible 
contributions from the light quarks, and this leads to their 
estimate $g_{hN}v_h\simeq 2m_N/9\simeq 210$ MeV. On the other hand,
recent result from lattice 
calculations\cite{lattice1,lattice2,lattice3,lattice4,lattice5,lattice6}, 
chiral perturbation theory\cite{strange1} and sum rules\cite{strange2} 
indicate $0\le y_N\le 0.1$ and $\sigma_{\pi N}\le 55$ MeV. 
This yields an estimate for the possible
range of the effective Higgs-nucleon coupling,
\begin{equation}\label{eq:gvh5}
210\,\mathrm{MeV}\lesssim g_{hN}v_h\lesssim 310\,\mathrm{MeV},
\end{equation}
and amounts to a factor of 2.2 of uncertainty in the nuclear recoil cross 
sections of Higgs portal dark matter particles. Furthermore, 
Durr \textit{et al.} find 
a higher strangeness content\cite{lattice7} $y_N=0.20(8)$ with a  
lower $\sigma_{\pi N}=38(3)(3)$ MeV, where the errors are statistical and systematic.
The large $y_N$ results from a large $\sigma_{sN}=105(41)(37)$ MeV, and this translates 
into a large effective Higgs-nucleon coupling 
(all errors added in quadrature) $g_{hN}v_h=320(43)$ MeV.
Hoferichter\cite{hoferichter} \textit{et al.} find a coupling equivalent to
$g_{hN}v_h=289(17)$ MeV from averaging $f_q^N$ values and estimating related
errors from four of the recent lattice calculations\cite{lattice4,lattice5,lattice6,lattice7}. 
However, the lattice calculations themselves do not have mutually overlapping error bars,
e.g. the results of Ref. \refcite{lattice6} yield $g_{hN}v_h=263(10)$ MeV.
Therefore, we are still more comfortable with the use of a possible 
range of values for $g_{hN}v_h$.
We also note that the recent evaluations by ATLAS\cite{atlasg} and CMS\cite{cms} 
used large uncertainties of $0.26\le f_N\le 0.66$ and $0.260\le f_N\le 0.629$, 
respectively, and that the careful evaluation 
by Alarc\'on \textit{et al.} from pion-nucleon scattering and pionic atomic 
spectroscopy\cite{alarcon} yields $\sigma_{\pi N}=59(7)$ MeV 
and $g_{hN}v_h\simeq 263(66)$ MeV, in good agreement with the conservative 
range (\ref{eq:gvh5}).

We therefore use the SVZ reference point $g_{hN}v_h=210$ MeV for displaying limits from
nucleon recoil cross sections for Higgs portal matter, because these 
provide the least stringent constraints while including 
the allowed mass ranges for\cite{lattice6,alarcon} $g_{hN}v_h=263$ MeV 
and for the Hoferichter \textit{et al.}
value $g_{hN}v_h=289$ MeV. This is not 
borne out of a desire to protect any of the testable Higgs portal models from 
early elimination, but out of caution. Constraining 
dark matter models from absence of a signal is different from dark matter mass 
reconstruction from a signal, and should be based on the least constraining 
available parameter estimates. However, we also report mass limits for
an effective Higgs-nucleon coupling $g_{hN}v_h=289$ MeV.

\section{Scalar Higgs Portal Dark Matter}	
\label{sec:scalar}

The coupling $g_S S^2 H^+H$ of a scalar electroweak singlet $S$
to the Higgs field $H$ in unitary gauge
\[
H=\left(\begin{array}{c} \phi^+ \\ \phi^0\\ \end{array}\right)
\rightarrow\frac{v_h+h}{\sqrt{2}}\left(\begin{array}{c} 0\\ 1\\ \end{array}\right),
\]
yields the minimal renormalizable dark matter addition $\mathcal{L}_S$ to the
Lagrangian of the Standard Model,
\begin{equation} \label{eq:LS}
\mathcal{L}_S= -\,\frac{1}{2}\partial S\cdot\partial S
-\frac{1}{2}m_S^2 S^2-\frac{\lambda_S}{4}S^4
-g_S v_h S^2 h-\frac{g_S}{2} S^2 h^2.
\end{equation}

This minimal Standard Model extension has been suggested on numerous 
occasions as a dark matter model\cite{zee,bento,cliff,dklm,wells,frank}
or as a complementary dark sector component\cite{cnw,kusenko}. The very 
small parameter space provides this model with very high predictive
power, and implications for indirect dark matter 
signals\cite{mcdonald,rd1,yaguna1,yaguna2,yaguna3,MDMind4,mike,rdfs1,FPU,beniwal,escudero} 
and direct signals\cite{barger1,yaguna3,djouadi1,frigero,bazzo,cline,strumia1,rdfs,sibo1,
duerr,ghosh,beniwal,sibo2,jusak,sibo,FPU,escudero,casas,gambit,prof2} has been the subject 
of numerous investigations. The stability poperties of the model and its 
extension to a complex scalar have been studied in 
Refs.~\refcite{gonderinger1,profumo,gonderinger2,lebedev1,petcov,khoze,khan,lebedev2,garg},
and have also been studied as a consequence of radiative conformal symmetry 
breaking in Refs.~\refcite{tom,mariana}.

Analysis of the correlation between dark matter mass and couplings
from thermal creation requires the corresponding dark matter
annihilation cross sections. For completeness we recall the leading
order contributions for scalar singlet annihilations into 
Higgs particles, fermions, and gauge bosons\cite{mcdonald},
\begin{equation} \label{eq:SS2hh}
\sigma_{SS\to hh}(s)=\frac{g_S^2\sqrt{s-4m_h^2}}{
8\pi s\sqrt{s-4m_S^2}}\left(\frac{s+2m_h^2}{s-m_h^2}\right)^2,
\end{equation}
\begin{equation} \label{eq:SS2ff}
\sigma_{SS\to f\overline{f}}(s)
=N_cg_S^2\frac{\sqrt{s-4m_f^2}^3}{
2\pi s\sqrt{s-4m_S^2}}
\frac{m_f^2}{(s-m_h^2)^2+m_h^2\Gamma_h^2},
\end{equation}
with $N_c=1$ for leptons and $N_c=3$ for quarks, and
\begin{equation} \label{eq:SS2VV}
\sigma_{SS\to ZZ,W^+W^-}(s)
=\frac{g_S^2\sqrt{s-4m_{W,Z}^2}
}{4\pi s\sqrt{s-4m_S^2}(1+\delta_Z)}
\frac{(s-2m_{W,Z}^2)^2+8m_{W,Z}^4}{
(s-m_h^2)^2+m_h^2\Gamma_h^2}.
\end{equation}
Here $\delta_Z=1$ for annihilation into $Z$ bosons and $\delta_Z=0$ for
annihilation into $W^+ W^-$. The velocity weighted cross sections 
are $v\sigma=2\sqrt{1-(4m_S^2/s)}\sigma(s)$, and thermal averaging
is performed according to (\ref{eq:thermalGG}).

\begin{figure}[hptb]
\centerline{\psfig{file=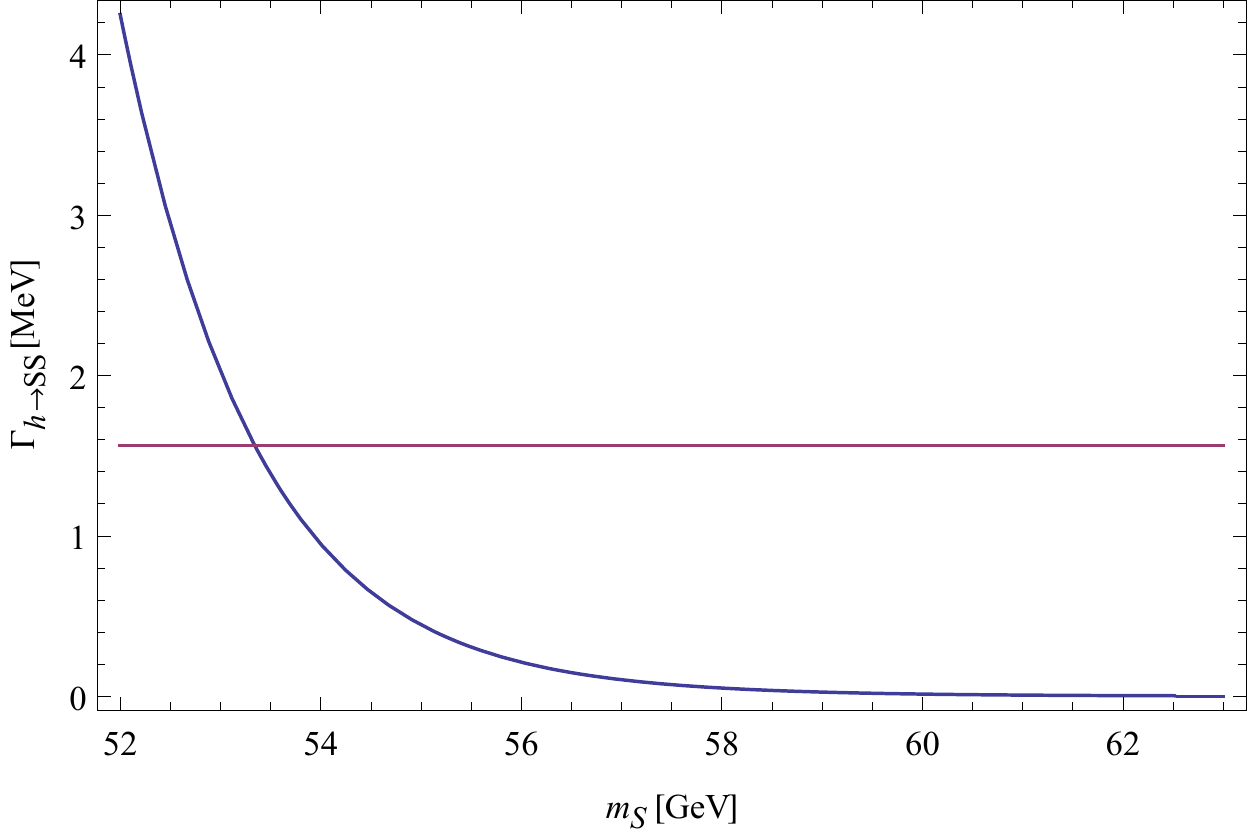,width=8cm}}
\vspace*{8pt}
\caption{The invisible decay width $\Gamma_{h\to SS}$ for $m_S$
between 52 GeV and $m_h/2$. The horizontal line arises from the limit on the 
branching ratio for invisible Higgs 
decays, $\mathcal{B}=\Gamma_{h\to\mathrm{inv.}}/(\Gamma_{h\to\mathrm{inv.}}
+\Gamma_{h\to\mathrm{SM}})\le 0.24$. The Higgs portal coupling $g_{hS}(m_S)$ is 
determined from the requirement that scalar Higgs portal matter accounts for 
the observed dark matter. \label{fig:h2ss}}
\end{figure}

Higgs portal dark matter from thermal freeze-out is constrained by 
direct search experiments, and in the low-mass 
sector $m_S\lesssim m_h/2\simeq 62.5$ GeV it is also constrained by the 
limits from the ATLAS and CMS collaborations\cite{cms,atlas} on the branching ratio
into invisible Higgs decays\cite{barger1,fox1,djouadi2,cline,atlas1}, see also
Ref.~\refcite{felix} for a general recent discussion of LHC dark matter searches.
These limits 
directly constrain the mass $m_S$ of Higgs portal dark matter, because the requirement
of the correct abundance $\varrho_D$ of the dark matter particles relates
dark matter coupling and mass, $g_S=f(m_S)$. The Higgs decay width
\begin{equation}\label{eq:h2SS}
\Gamma_{h\to SS}=\frac{g_S^2 v_h^2}{8\pi m_h^2}
\sqrt{m_h^2-4m_S^2}
\end{equation}
then implies the constraint $m_S\gtrsim 53.3$ GeV on scalar 
Higgs portal dark matter under the 
constraint\cite{cms} $\mathcal{B}=\Gamma_{h\to\mathrm{inv.}}/(\Gamma_{h\to\mathrm{inv.}}
+\Gamma_{h\to\mathrm{SM}})\le 0.24$ on the branching ratio for invisible 
Higgs decays, see Fig.~\ref{fig:h2ss}.

Note that this constraint becomes stronger if Higgs portal matter is not the
dominant dark matter component in the universe, because the required coupling
strength for creating a remnant dark matter density $\varrho_S$ from freeze-out
of particle annihilation scales like $g_S^2\propto \varrho_S^{-1}$. Larger
coupling implies later freeze-out and therefore smaller remnant abundance,
but also a larger contribution to the Higgs decay width. Assuming e.g.~that
remnant scalar Higgs portal matter contributes about 50\% of the dark matter
in the universe would increase the mass constraint to $m_S\gtrsim 54$ GeV.
On the other hand, if the dark matter particles can annihilate also through 
alternative channels besides virtual Higgs exchange,
then the Higgs portal coupling $g_S^2$ must only account for a 
smaller fraction $\zeta_{hS}\langle\sigma v\rangle_f$, $0\le\zeta_{hS}<1$,
of the required annihilation cross 
section $\langle\sigma v\rangle_f$ for thermal freeze-out, and this reduces
the coupling constraint $g_S^2=f(m_S)$ to $g_S^2=\zeta_{hS}f(m_S)$, thus also
reducing the mass constraint due to the reduced invisible Higgs decay 
width $\Gamma_{h\to SS}\to\zeta_{hS}\Gamma_{h\to SS}$. Therefore light dark matter
with a Higgs portal is still compatible with invisible Higgs decay constraints
if the Higgs portal is not the only connection between the dark matter 
and the baryonic sector.

The nucleon recoil cross section for perturbatively coupled bosonic Higgs portal
matter of mass $m_D$ and coupling $g_D$ ($D\in\{S,V\}$ for scalar or vector dark matter,
respectively) is
\begin{equation}\label{eq:sigmaSN}
\sigma_{DN}=\frac{g_{hN}^2 g_D^2 v_h^2}{\pi m_h^4}
\frac{m_N^2}{(m_D+m_N)^2}.
\end{equation}
Here $m_N$ is the nucleon mass and $g_{hN}$ is the coupling constant
in the effective Higgs-nucleon coupling term $g_{hN}\overline{N}Nh$.
While the derivation of this equation is more complicated for the vector
model due to the presence of helicity factors, averaging and summation 
over initial and final state helicities yields the same result as for
scalar dark matter. 

We use an effective nucleon mass of $m_N=930.6$ MeV for evaluations 
of (\ref{eq:sigmaSN}) and the corresponding formulas for fermionic Higgs portal
models discussed in the following sections, because this corresponds to the 
average nucleon mass in stable or long lived Xenon isotopes. However, our results 
can also be used for DEAP-3600\cite{deap} and DarkSide-20k\cite{darkside}. 
The average nucleon mass in stable Argon isotopes is 930.4 MeV. The relative error of 
order $2\times 10^{-4}$ is negligible for current purposes of comparing dark matter 
models to direct search limits. 

\begin{figure}[hptb]
\centerline{\psfig{file=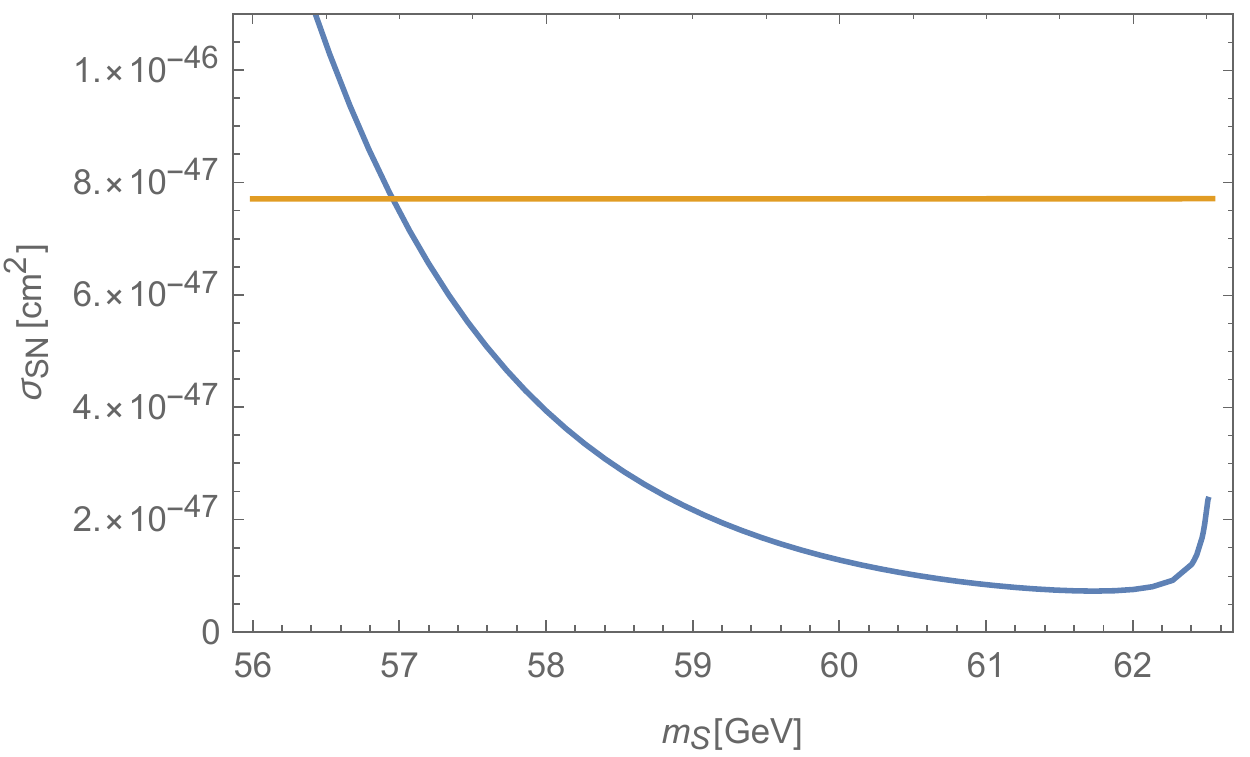,width=8cm}}
\vspace*{8pt}
\caption{The nucleon recoil cross section for scalar Higgs portal
dark matter in the resonance region $m_S\lesssim m_h/2$ for $g_{hN}v_h=210$ MeV.
The seemingly horizontal line from XENON1T is actually upwards curved,
but varies by less than 1\permil\ in the displayed mass 
range.\label{fig:sigmaSNlite}}
\end{figure}

The required small coupling constant $g_S$ near the resonance region $m_S\lesssim m_h/2$
implies a small recoil cross section in that region which is still compatible with
the cross section limits from XENON1T\cite{xenon}, which are comparable to but stronger
than the limits from PandaX-II\cite{panda} and LUX\cite{lux}.
In the resonance region, the mass range $57\,\mathrm{GeV}\le m_S<m_h/2$ 
($58\,\mathrm{GeV}\le m_S<m_h/2$ for $g_{hN}v_h=289$ MeV) complies with current
direct search limits, see Fig.~\ref{fig:sigmaSNlite}. 

The minimal recoil cross section near $m_S\simeq 62$ GeV
is $\sigma_{SN}\simeq 7.3\times 10^{-48}\,\mathrm{cm}^2$. 
Recoil cross sections which are that small can be tested with XENONnT\cite{xenon2}, 
LZ\cite{LZ}, DarkSide-20k\cite{darkside} or DARWIN\cite{darwin}.

\begin{figure}[hptb]
\centerline{\psfig{file=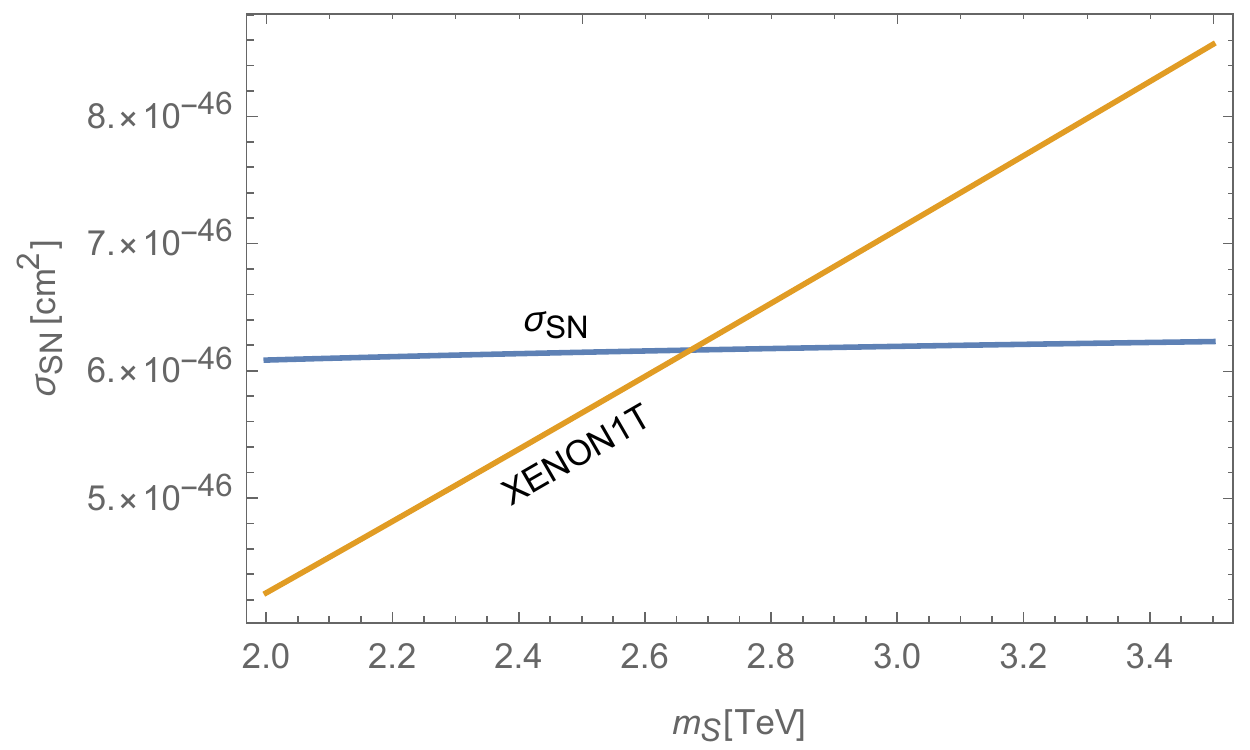,width=6cm}
\hspace{3mm}\psfig{file=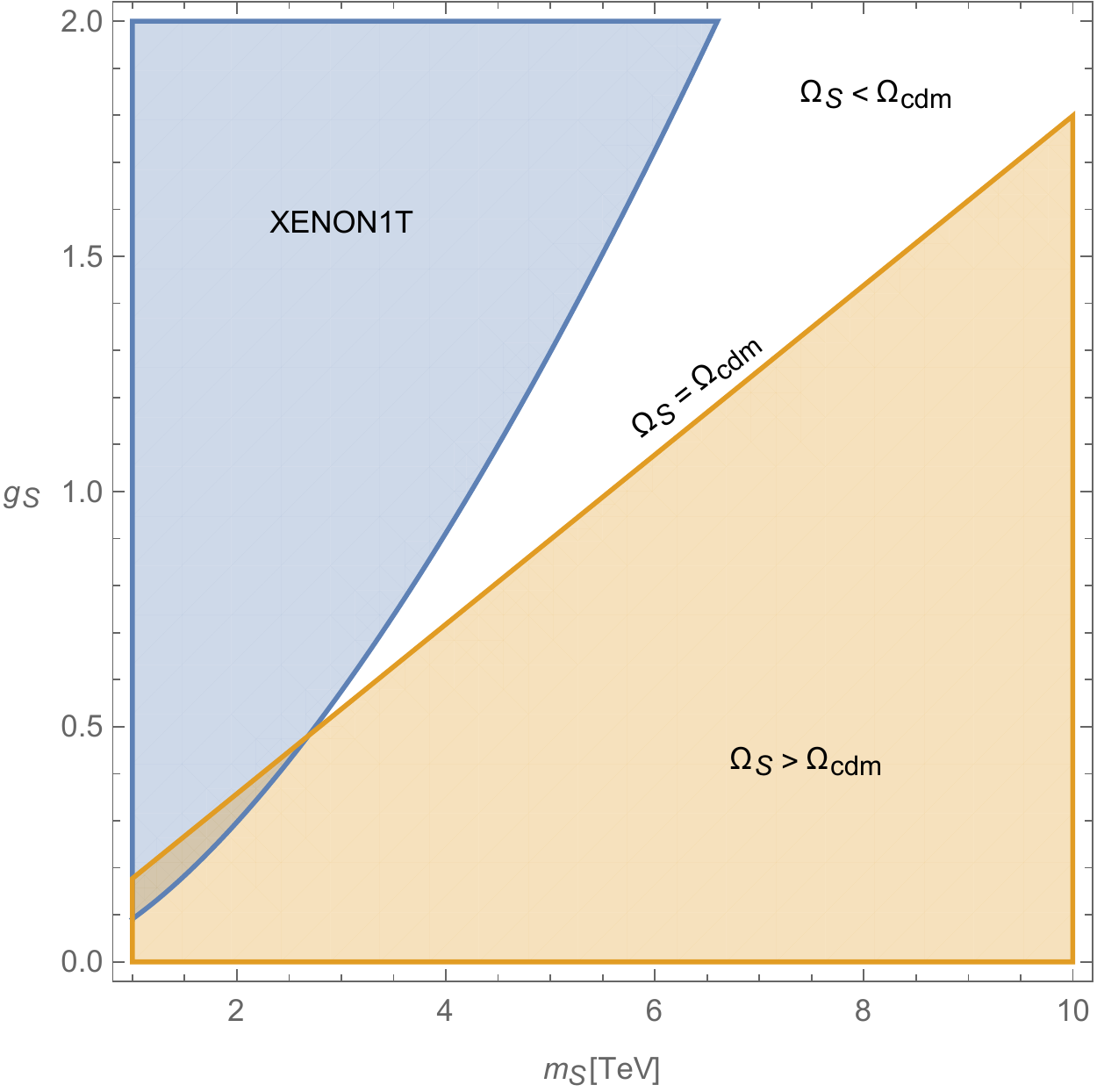,width=6cm}}
\vspace*{8pt}
\caption{Left panel:
The nucleon recoil cross section for scalar Higgs portal
dark matter with $\Omega_S=\Omega_{\mathrm{cdm}}$
in the mass range $2\,\mathrm{TeV}\le m_S\le 3.5\,\mathrm{TeV}$ versus
the extrapolated limit from XENON1T.
Right panel: Exclusion regions from XENON1T and the requirement
$\Omega_S\le\Omega_{\mathrm{cdm}}$.\label{fig:sigmaSNX}}
\end{figure}

In the high mass sector, extrapolation of the published results from XENON1T 
rules out scalar Higgs portal dark matter below $m_S\lesssim 2.7$ TeV for
$g_{hN}v_h=210$ MeV, see Fig.~\ref{fig:sigmaSNX}.
This is also often displayed in the form of an exclusion region in 
the $(m_S,g_S)$-plane, see the right panel in Fig.~\ref{fig:sigmaSNX}, 
where the yellow region is excluded by the 
requirement $\Omega_S\le\Omega_{\mathrm{cdm}}$. The excluded mass range above $m_h/2$ 
increases to $m_S\lesssim 4.5$ TeV for stronger Higgs-nucleon 
coupling $g_{hN}v_h=289$ MeV.

\begin{figure}[hpb]
\centerline{\psfig{file=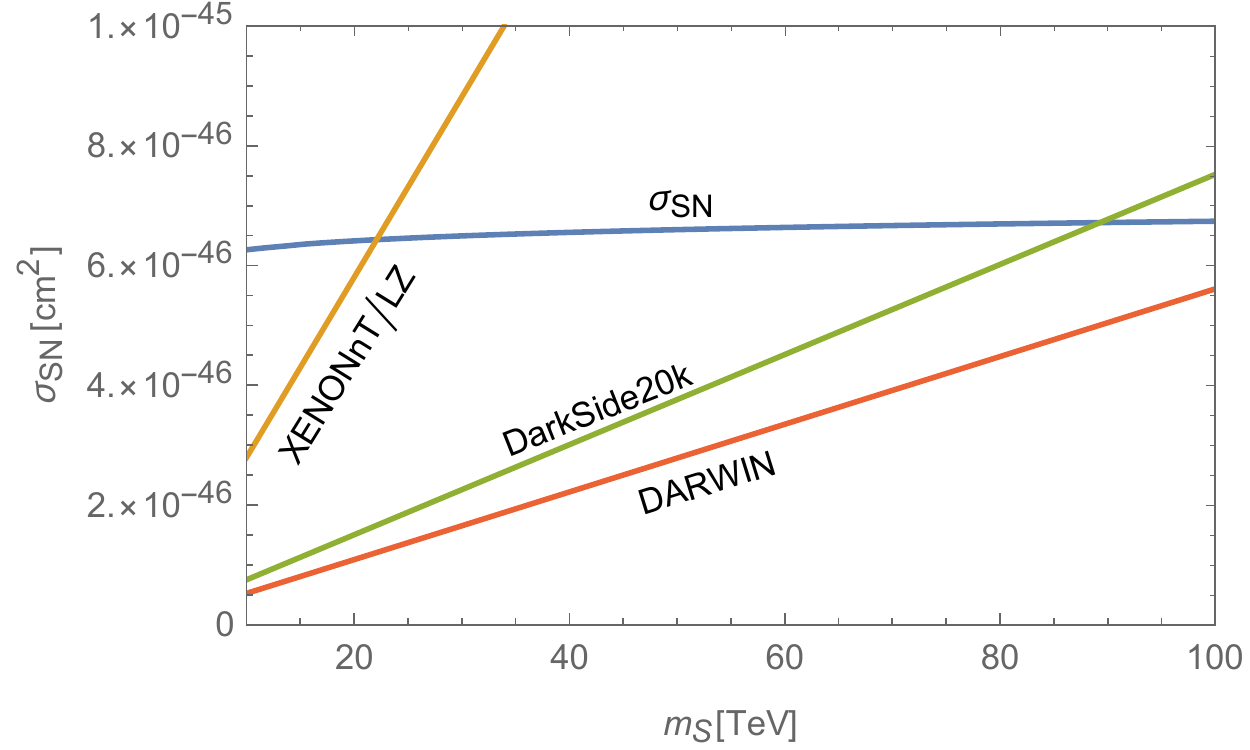,width=8cm}}
\vspace*{8pt}
\caption{The estimated nucleon recoil cross section for scalar Higgs portal
dark matter in the mass range $10\,\mathrm{TeV}\le m_S\le 100\,\mathrm{TeV}$ versus
extrapolated sensitivities for XENONnT, LZ, DarkSide-20k and DARWIN. \label{fig:sigmaSND}}
\end{figure}

In the very high mass region, we can use the perturbative formula (\ref{eq:sigmaSN})
as well as the annihilation cross sections (\ref{eq:SS2hh})-(\ref{eq:SS2VV}) 
only up to a maximal mass value $m_{lD}$ when the dark matter coupling
approaches the perturbativity limit $g_D\lesssim 4\pi$. For the scalar model,
this limit is reached for $m_{lS}\lesssim 67$ TeV. Beyond this mass value,
the scalar Higgs portal recoil cross section in Fig.~\ref{fig:sigmaSND}
should only be considered as an order of magnitude estimate.
It is nevertheless intriguing that XENONnT and LZ might test scalar Higgs 
portal dark matter up to about 20 TeV, while DarkSide-20k and DARWIN with 
200 ton-year exposures could potentially cover the full mass range for frozen-out
WIMPS up to the unitarity limit, see Fig.~\ref{fig:sigmaSND}.

\section{Vector Higgs Portal Dark Matter}	
\label{sec:vector}

Vector Higgs portal dark matter
\begin{equation}\label{eq:LV}
\mathcal{L}_V=-\,\frac{1}{4}V_{\mu\nu}V^{\mu\nu}
-\frac{1}{2}m_V^2 V_\mu V^\mu -\frac{\lambda_V}{4}(V_\mu V^\mu)^2 
-g_V v_h V_\mu V^\mu h-\frac{g_V}{2}V_\mu V^\mu h^2,
\end{equation}
\begin{equation} \label{eq:LFV}
V_{\mu\nu}\equiv\partial_\mu V_\nu-\partial_\nu V_\mu,
\end{equation}
can arise in a renormalizable model from spontaneous symmetry breaking
in the dark sector\cite{V1} and has been further discussed also in 
Refs.~\refcite{djouadi1,djouadi2,baek,rdfs,gross,duch,V2,jusak1,prof2,escudero}.
The cross sections for annihilation of the dark vector bosons are
\begin{equation} \label{eq:VV2hh}
\sigma_{VV\to hh}=\frac{g_V^2\sqrt{s-4m_h^2}}{288\pi s\sqrt{s-4m_V^2}}
\left(\frac{s+2m_h^2}{s-m_h^2}\right)^2
\frac{(s-2m_{V}^2)^2+8m_{V}^4}{m_{V}^4},
\end{equation}
\begin{equation} \label{eq:VV2ff}
\sigma_{VV\to f\overline{f}}=N_c\frac{g_V^2 m_f^2\sqrt{s-4m_f^2}^3}{
72\pi m_{V}^4 s\sqrt{s-4m_V^2}}
\frac{(s-2m_{V}^2)^2+8m_{V}^4}{(s-m_h^2)^2+m_h^2\Gamma_h^2},
\end{equation}
and
\begin{eqnarray}\nonumber
\sigma_{VV\to ZZ,W^+ W^-}&=&
\frac{g_V^2\sqrt{s-4m_{W,Z}^2}}{144\pi(1+\delta_Z) s\sqrt{s-4m_V^2}}
\frac{(s-2m_{W,Z}^2)^2+8m_{W,Z}^4}{(s-m_h^2)^2+m_h^2\Gamma_h^2}
\\ \label{eq:VV2VV}
&&\times
\frac{(s-2m_{V}^2)^2+8m_{V}^4}{m_{V}^4},
\end{eqnarray}
These low-energy effective annihilation cross sections for $s\lesssim 8m_V^2$
do not satisfy the unitarity constraint of bounded $\lim_{s\to\infty}s\sigma(s)$, 
which is a familiar indication of the need of UV completion of the vector Higgs 
portal model through spontaneous symmetry breaking in the dark 
sector\cite{V1,baek,duch}. For the same reason, the $m_V{}^{-4}$ term in the 
denominators of 
Eqs.~(\ref{eq:VV2hh}-\ref{eq:VV2VV})
(and also in Eq.~(\ref{eq:h2VV}) below) seems to appear unphysical, but is
a well-known consequence of averaging over three massive (instead
of two massless) polarization states\cite{rdfs,gross,duch,prof2}.
We can use polarization vectors for the choice $\bm{e}_z\|\bm{p}$,
\begin{eqnarray}\nonumber
\epsilon^{(1)}(\bm{p})&=&(0,1,0,0),\quad
\epsilon^{(2)}(\bm{p})=(0,0,1,0),
\\ \label{eq:polstatesV}
\epsilon^{(3)}(\bm{p})&=&
\frac{1}{m_{V}}(|\bm{p}|,0,0,\sqrt{\bm{p}^2+m_{V}^2}).
\end{eqnarray}
The sum over the initial polarization states yields
\[
\sum_\alpha\epsilon^{(\alpha)}(\bm{p})\otimes\epsilon^{(\alpha)}(\bm{p})
=\left(\begin{array}{cccc}
\bm{p}^2/m_{V}^2 &\,\, 0 &\,\, 0 &\,\, 
|\bm{p}|\sqrt{\bm{p}^2+m_{V}^2}/m_{V}^2\\
0 &\,\, 1 &\,\, 0 &\,\, 0 \\
0 &\,\, 0 &\,\, 1 &\,\, 0 \\
|\bm{p}|\sqrt{\bm{p}^2+m_{V}{}^2}/m_{V}^2&\,\,
0 &\,\, 0 &\,\, (\bm{p}^2+m_{V}{}^2)/m_{V}^2\\
\end{array}\right),
\]
i.e.
\begin{equation}\label{eq:polstatesum1}
\sum_\alpha\epsilon^{(\alpha)\mu}(\bm{p})
\otimes\epsilon^{(\alpha)\nu}(\bm{p})
=\eta^{\mu\nu}+\frac{p^\mu p^\nu}{m_{V}^2}.
\end{equation}
This implies
\begin{equation}\label{eq:polstatesum2}
\sum_{\alpha,\beta}\left(\epsilon^{(\alpha)}(\bm{p}_1)
\cdot\epsilon^{(\beta)}(\bm{p}_2)\right)^2
=\left(\eta^{\mu\nu}+\frac{p_1^\mu p_1^\nu}{m_{V}^2}\right)
\left(\eta_{\nu\mu}+\frac{p_{2\nu} p_{2\mu}}{m_{V}^2}\right)
=2+\frac{(p_1\cdot p_2)^2}{m_{V}^4}.
\end{equation}
However, the low energy annihilation cross 
sections\cite{V1,baek,rdfs,gross,duch,prof2} (\ref{eq:VV2hh}-\ref{eq:VV2VV})
can be used for the analysis of cosmological implications and constraints of 
vector Higgs portal models because thermal averaging (\ref{eq:thermalGG}) cuts off 
the high energy parts with $\exp(-\sqrt{s}/T)\lesssim\exp(-25\sqrt{s}/m_V)$
near $T=T_f$.

\begin{figure}[hptb]
\centerline{\psfig{file=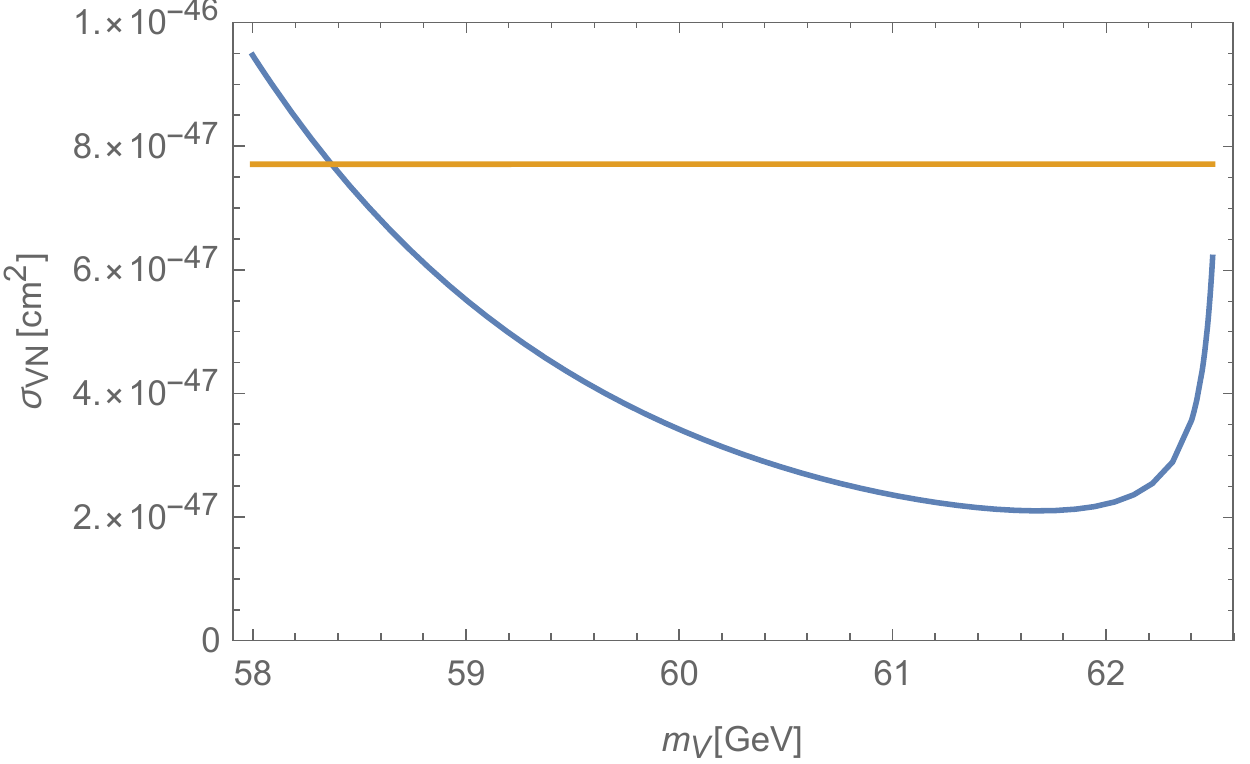,width=8cm}}
\vspace*{8pt}
\caption{The nucleon recoil cross section for vector Higgs portal
dark matter in the resonance region $m_V\lesssim m_h/2$ versus the limits from
XENON1T. \label{fig:sigmaVN}}
\end{figure}

The contribution to the Higgs decay width
\begin{equation} \label{eq:h2VV}
\Gamma_{h\to VV}=\frac{g_V^2 v_h^2}{32\pi m_h^2}\sqrt{m_h^2-4m_V^2}
\frac{(m_h^2-2m_{V}^2)^2+8m_{V}^4}{m_{V}^4},
\end{equation}
implies the constraint $m_V\gtrsim 56.3$ GeV on vector Higgs portal dark matter
using again the limit $\mathcal{B}\le 0.24$ on the branching ratio into invisible 
Higgs decays, cf. Fig.~\ref{fig:h2ss} for the corresponding constraint for the minimal
scalar Higgs portal dark matter.
However, just like in the scalar case, the direct search limit is also stronger for
vector Higgs portal dark matter.
The required small coupling constant $g_V$ near the resonance region $m_V\lesssim m_h/2$
implies a small recoil cross section in that region which is so far still compatible with
the direct search constraints\cite{xenon} for $58.4\,\mathrm{GeV}< m_V<m_h/2$
if $g_{hN}v_h=210$ MeV, see Fig.~\ref{fig:sigmaVN}. A stronger Higgs-nucleon coupling
of $g_{hN}v_h=289$ MeV decreases this mass range to $59.6\,\mathrm{GeV}< m_V<m_h/2$.
The minimal recoil cross section near $m_V\simeq 62$ GeV,
 $\sigma_{VN}\simeq 2.1\times 10^{-47}\,\mathrm{cm}^2$, will be within reach 
of XENONnT, LZ, DarkSide-20k and DARWIN.

\begin{figure}[hptb]
\centerline{\psfig{file=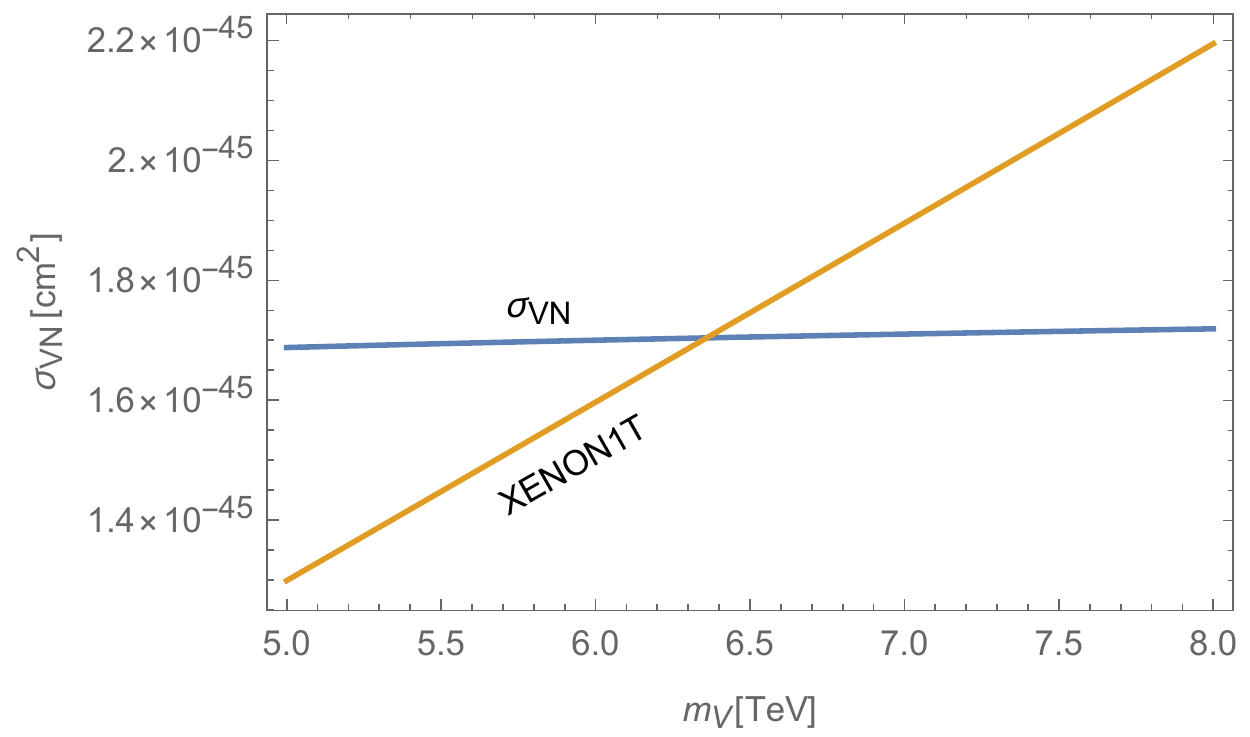,width=6cm}\hspace{3mm}
\psfig{file=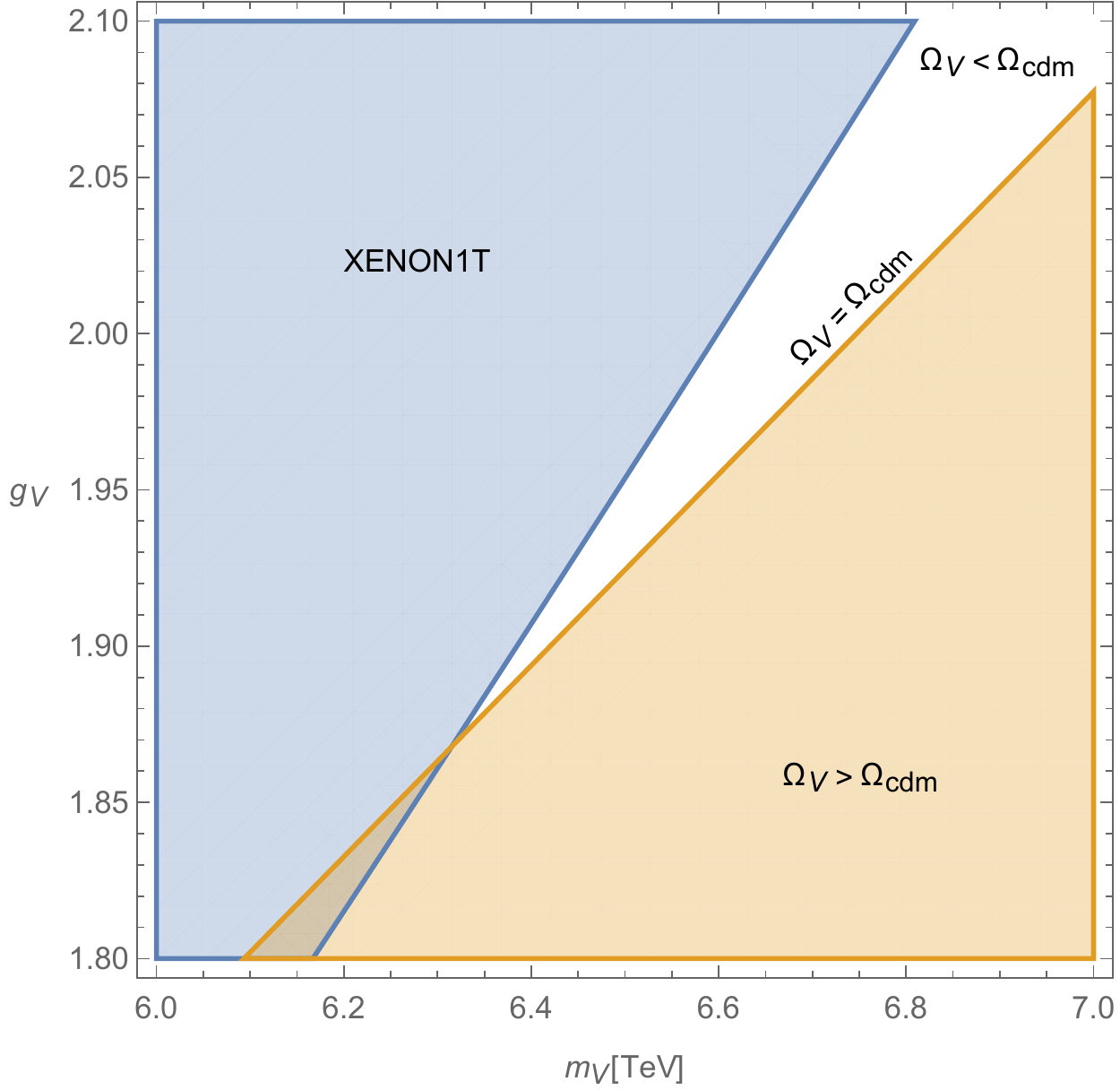,width=6cm}}
\vspace*{8pt}
\caption{Left panel: The nucleon recoil cross section for vector Higgs portal
dark matter with $\Omega_V=\Omega_{\mathrm{cdm}}$ and $g_{hN}v_h=210$ MeV
in the mass range $5\,\mathrm{TeV}\le m_V\le 8\,\mathrm{TeV}$ versus
the extrapolated limit from XENON1T. Right panel: Exclusion regions in the
$(m_V,g_V)$ plane from XENON1T and the 
requirement $\Omega_V\le\Omega_{\mathrm{cdm}}$.\label{fig:VNTeV}}
\end{figure}

In the high mass sector, extrapolation of the published results from XENON1T rules out vector 
Higgs portal dark matter below $m_V\lesssim 6.4$ TeV
(or below $m_V\lesssim 11.7$ TeV for $g_{hN}v_h=289$ MeV), 
see Fig.~\ref{fig:VNTeV}.

\begin{figure}[hptb]
\centerline{\psfig{file=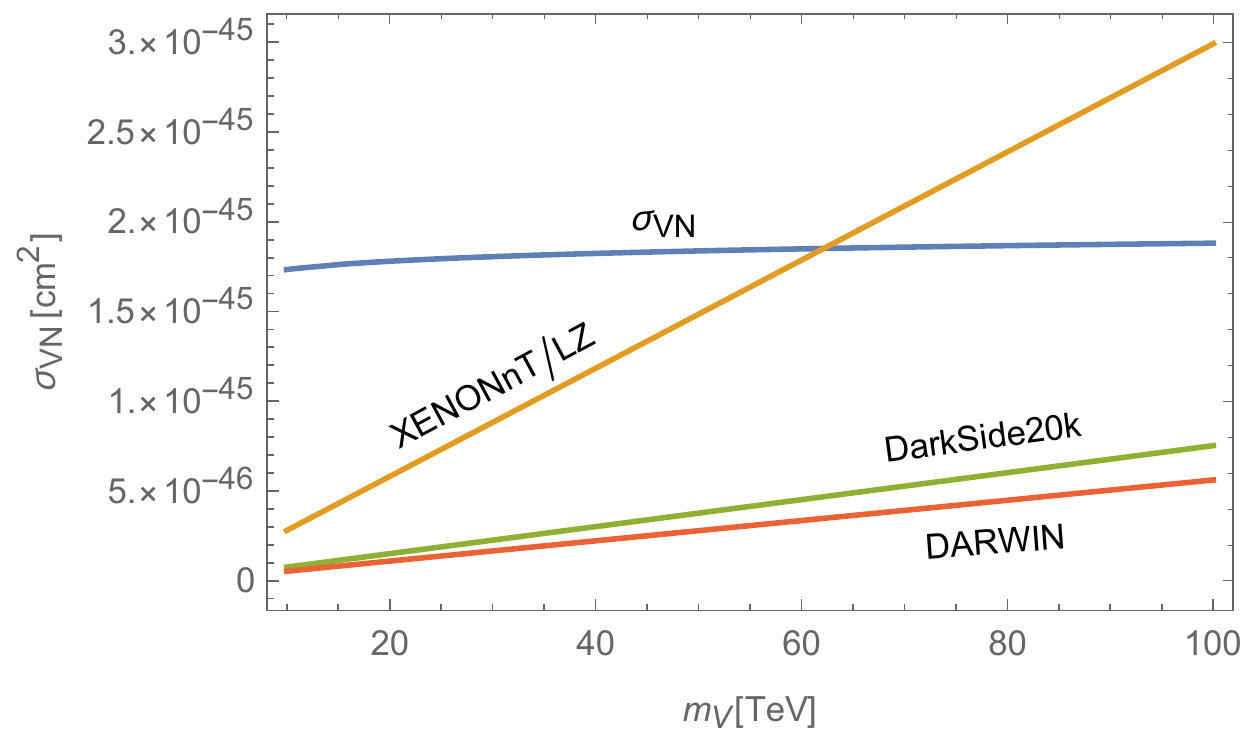,width=8cm}}
\vspace*{8pt}
\caption{The nucleon recoil cross section for vector Higgs portal
dark matter in the mass range $10\,\mathrm{TeV}\le m_V\le 100\,\mathrm{TeV}$ versus
the extrapolated sensitivities for XENONnT, LZ, DarkSide-20k and DARWIN. \label{fig:VNheavy}}
\end{figure}

 For yet higher masses, we have to caution that the coupling will reach the perturbativity
limit $g_V\lesssim 4\pi$ for a mass limit $m_{lV}\lesssim 40$ TeV. However, with the caveat that
the calculated recoil cross section should at best be considered as an order of magnitude
estimate beyond that limit, it is intriguing that XENONnT and LZ may already have the potential 
to cover the mass range for vector Higgs portal dark matter up to about 60 TeV if $g_{hN}v_h=210$ MeV, 
see Fig.~\ref{fig:VNheavy}. The possible exclusion region would reach the unitarity limit
at 100 TeV if $g_{hN}v_h=289$ MeV.

\section{Fermionic Higgs Portal Dark Matter}	
\label{sec:fermions}

Higgs portal couplings of the form 
\begin{equation}\label{eq:darkchi1}
\mathcal{H}=\frac{1}{M}\overline{\chi}\cdot\Gamma\cdot\chi\left(
H^+\cdot H-\frac{v_h^2}{2}\right)
\end{equation}
to dark fermions $\chi$ can be generated from Yukawa couplings to a scalar 
mediator $\phi$,
\[
\mathcal{H}_\phi=\frac{1}{2}m_\phi^2\phi^2+g\phi\overline{\chi}\cdot\Gamma\cdot\chi
+\lambda\phi\left(H^+\cdot H-\frac{v_h^2}{2}\right)
\]
with $M=-m_\phi^2/g\lambda$. The coupling scale $M$ 
satisfies $|M|<m_\phi$ if $m_\phi<|g\lambda|$, i.e. the coupling scale $M$ itself
does not necessarily set the scale for new additional degrees of freedom besides
the dark matter, and therefore cannot be used to infer a limit on the validity
range of the effective Higgs portal couplings (\ref{eq:darkchi1}). 
These kinds of fermionic Higgs portal models were discussed in 
Refs.~\refcite{djouadi1,strumia1,maxim1,kimlee,maxim2,maxim3,zupan,fedderke,escudero,jusak1,jusak2,prof2}. 
Thermal freeze-out determines the size of the effective Higgs portal 
coupling $v_h/M$ as a function of the dark fermion mass $m_\chi$. These models remain 
perturbative up to the unitarity limit, see 
e.g.~Figs.~\ref{fig:m56to63}, \ref{fig:m63to300}, \ref{fig:mu56to63} 
and \ref{fig:mu63to300} below.
  
\subsection{CP even coupling}

\begin{figure}[hptb]
\centerline{\psfig{file=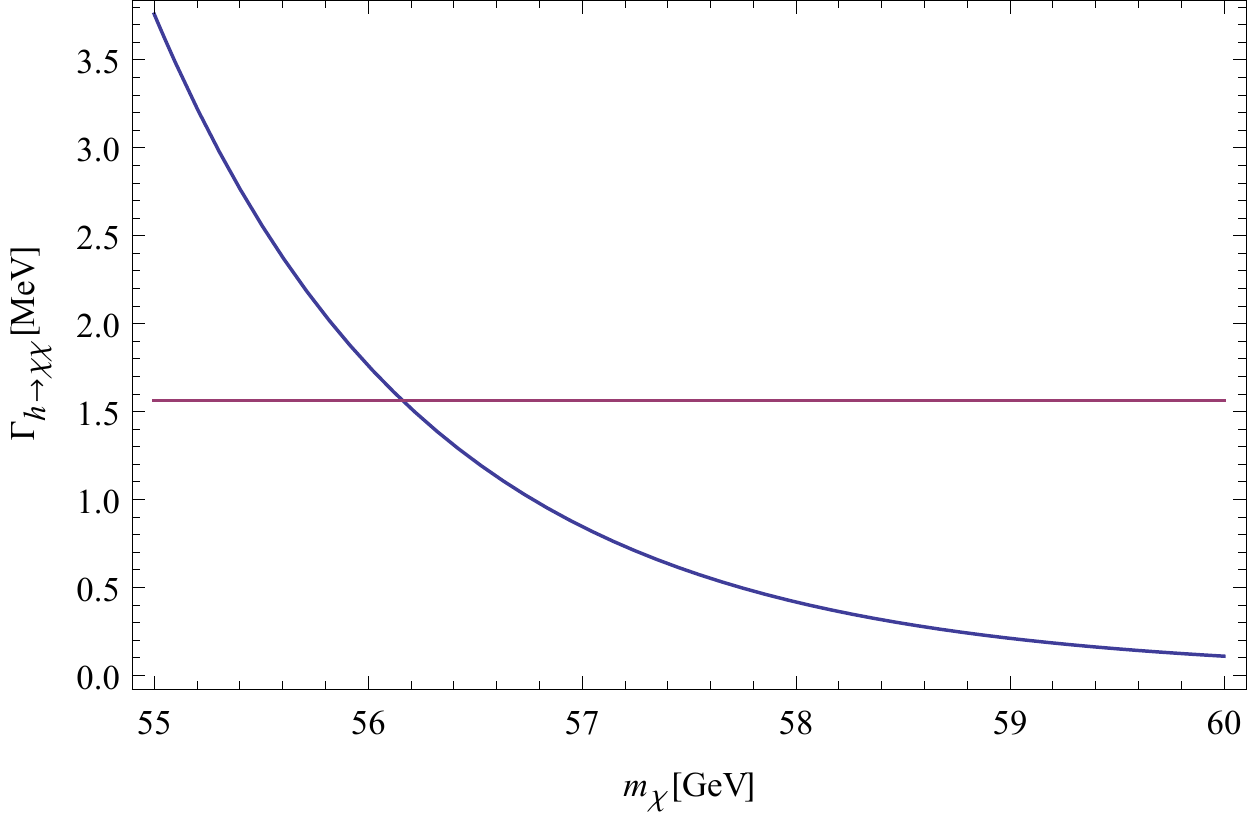,width=8cm}}
\vspace*{8pt}
\caption{The invisible decay width $\Gamma_{h\to\chi\overline{\chi}}$ for 
the CP even coupling (\ref{eq:Lchi}) and $m_\chi$
between 55 GeV and 60 GeV. The horizontal line arises from the limit on the 
branching ratio for invisible Higgs 
decays, $\mathcal{B}=\Gamma_{h\to\mathrm{inv.}}/(\Gamma_{h\to\mathrm{inv.}}
+\Gamma_{h\to\mathrm{SM}})\le 0.24$. The mass parameter $M$ for the $h\chi^2$ coupling 
is determined from the requirement that the fermionic Higgs portal matter accounts for 
the observed dark matter. \label{fig:h2chichi} }
\end{figure}

The CP even fermionic Higgs portal model
\begin{eqnarray}\nonumber
\mathcal{L}_\chi&=&
\overline{\chi}\left(\mathrm{i}\gamma^\mu\partial_\mu-m^{(0)}_\chi\right)\chi
-\frac{1}{M}\overline{\chi}\chi H^+\cdot H
\\ \label{eq:Lchi}
&=&\overline{\chi}\left(\mathrm{i}\gamma^\mu\partial_\mu-m_\chi\right)\chi
-\frac{v_h}{M} \overline{\chi}\chi h-\frac{1}{2M}\overline{\chi}\chi  h^2
\end{eqnarray}
yields annihilation cross sections
\begin{equation} \label{eq:chichi2hh}
\sigma_{\chi\overline{\chi}\to hh}(s)=
\frac{\sqrt{s-4m_\chi^2}\sqrt{s-4m_h^2}}{64\pi M^2 s}
\left(\frac{s+2m_h^2}{s-m_h^2}\right)^2,
\end{equation}
\begin{equation} \label{eq:chichi2ff}
\sigma_{\chi\overline{\chi}\to f\overline{f}}(s)
=N_c\frac{\sqrt{s-4m_\chi^2}\sqrt{s-4m_f^2}^3}{
16\pi M^2 s}\frac{m_f^2}{
(s-m_h^2)^2+m_h^2\Gamma_h^2},
\end{equation}
\begin{equation} \label{eq:chichi2VV}
\sigma_{\chi\overline{\chi}\to ZZ,W^+ W^-}(s)
=\frac{\sqrt{s-4m_\chi^2}\sqrt{s-4m_{W,Z}^2}}{32\pi M^2(1+\delta_Z)s}
\frac{(s-2m_{W,Z}^2)^2+8m_{W,Z}^4}{
(s-m_h^2)^2+m_h^2\Gamma_h^2}.
\end{equation}
The corresponding contribution to the Higgs decay width for $m_\chi<m_h/2$,
\begin{equation}\label{eq:h2chichi}
\Gamma_{h\to\chi\overline{\chi}}
=\frac{v_h^2}{8\pi M^2 m_h^2}\sqrt{m_h^2-4m_\chi^2}^3,
\end{equation}
implies the constraint $m_\chi\gtrsim 56.2$ GeV for CP even fermionic Higgs portal 
matter, see Fig.~\ref{fig:h2chichi}.

However, whereas for the bosonic models the current direct search constraints in the 
resonance region were comparable to the Higgs decay constraints, the recoil
cross sections of order $\sigma_{\chi N}\simeq 3\times 10^{-46}\,\mathrm{cm}^2$ for
the model (\ref{eq:Lchi}) in the region $m_\chi\sim m_h/2$
are already ruled out by the direct search experiments.

\begin{figure}[hptb]
\centerline{\psfig{file=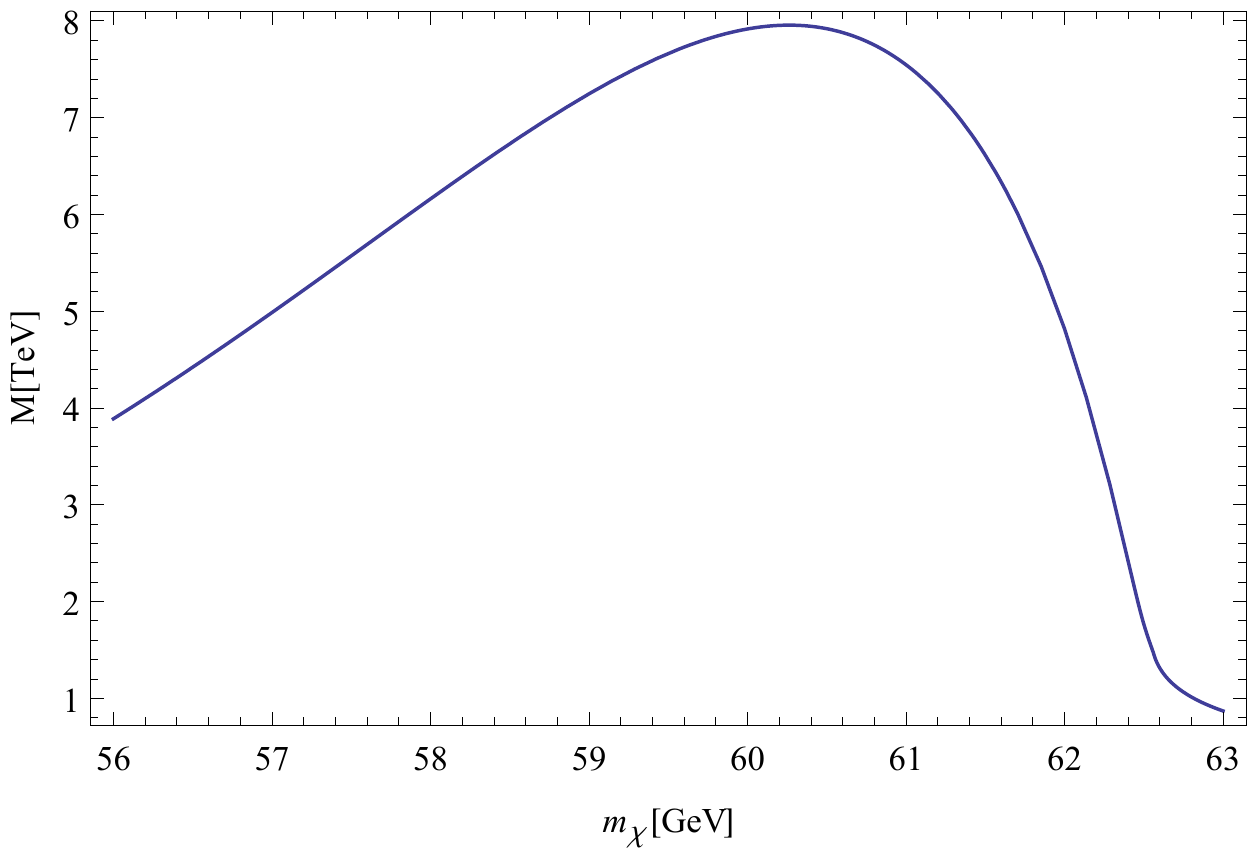,width=8cm}}
\vspace*{8pt}
\caption{The required coupling scale $M$ for dark matter creation through
the fermionic Higgs portal (\ref{eq:Lchi}) in the WIMP mass range $56\,\mathrm{GeV}
\le m_\chi\le 63\,\mathrm{GeV}$. \label{fig:m56to63} }
\end{figure}

The coupling scale $M$ can be determined from the requirement that thermal freeze-out
of the fermionic Higgs portal matter creates the observed dark matter abundance.
It varies in the 1-8 TeV range for dark matter masses in the 
range $56\,\mathrm{GeV}\le m_\chi\le 63\,\mathrm{GeV}$, see Fig.~\ref{fig:m56to63}.

\begin{figure}[hptb]
\centerline{\psfig{file=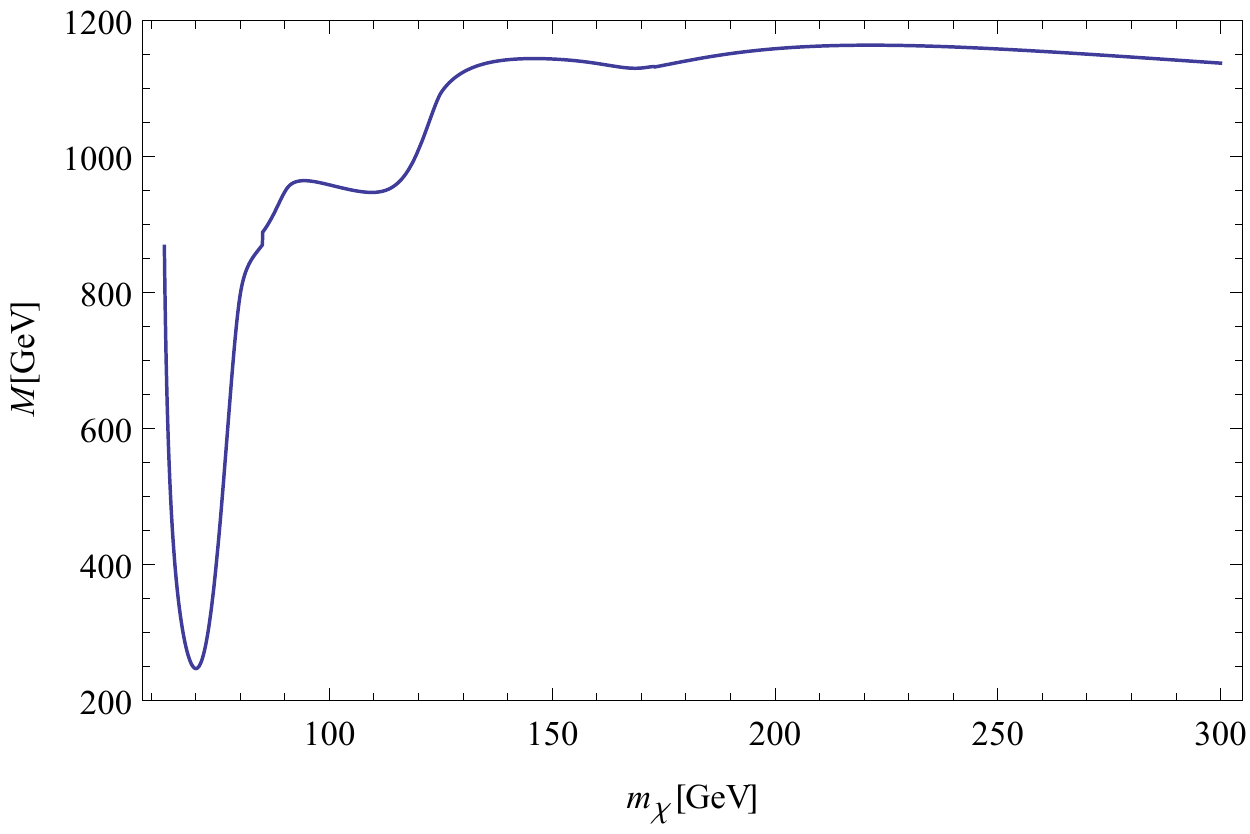,width=8cm}}
\vspace*{8pt}
\caption{The required coupling scale $M$ for dark matter creation through
the fermionic Higgs portal (\ref{eq:Lchi}) in the WIMP mass range $63\,\mathrm{GeV}
\le m_\chi\le 300\,\mathrm{GeV}$. \label{fig:m63to300} }
\end{figure}

The required scale $M$ increases near mass thresholds for $\chi\overline{\chi}$
annihilation, because opening up new annihilation channels implies that thermal freeze-out
can produce the observed dark matter abundance with smaller 
coupling $g_{h\chi}=v_h/M$, see Fig.~\ref{fig:m63to300}. 

Above the top mass, $M$ decreases logarithmically with increasing dark matter 
mass $m_\chi$ because the required cross section $\langle\sigma v\rangle(T)$ for 
thermal creation decreases with increasing $m_\chi$. The decrease in $M$ compensates for 
the decrease in $\langle\sigma v\rangle(T)$ to ensure that the 
requirement $\langle\sigma v\rangle(T_f)=\langle\sigma v\rangle_f\equiv
\langle\sigma v\rangle|_{\Omega_\chi=\Omega_{\mathrm{cdm}}}$
can still be met. The asymptotic value of $M$ near the unitarity 
limit $m_\chi\simeq 100$ TeV is $M\simeq 860$ GeV.

\begin{figure}[hptb]
\centerline{\psfig{file=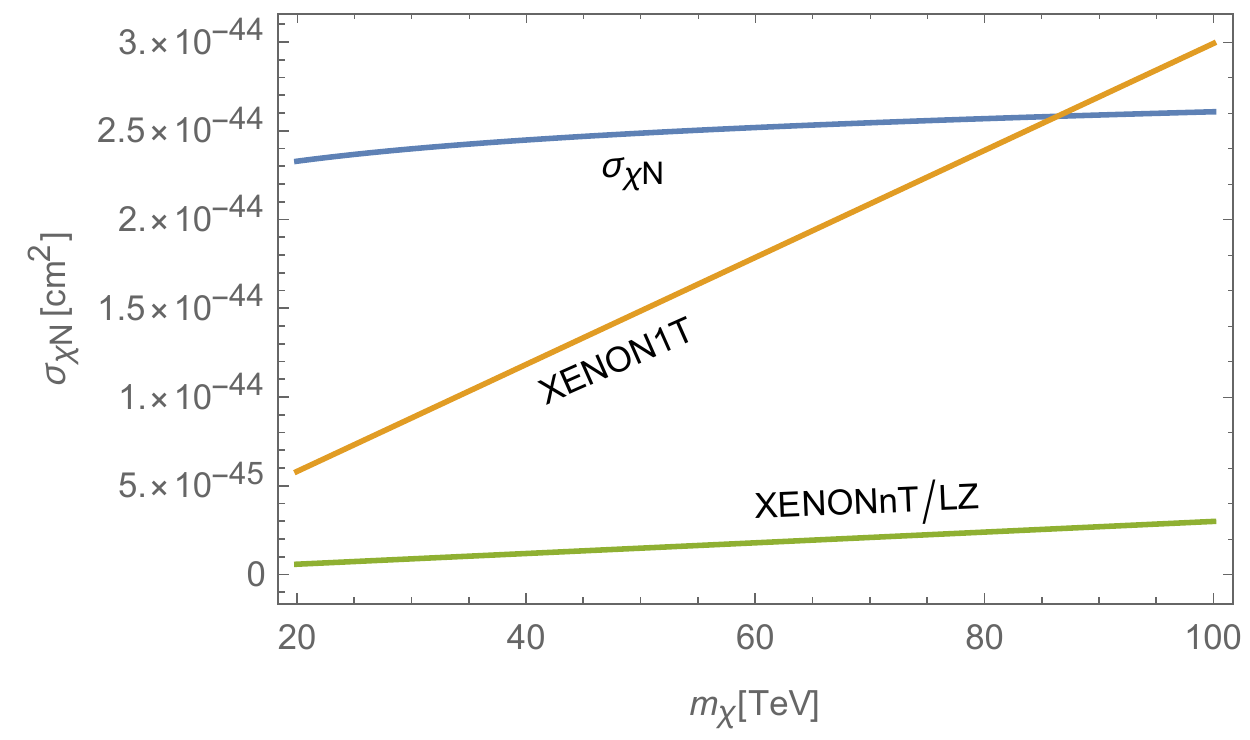,width=8cm}}
\vspace*{8pt}
\caption{The recoil cross section for fermionic Higgs portal
dark matter with CP even coupling (\ref{eq:Lchi}) versus the
extrapolated XENON1T limit. \label{fig:chiheavy}}
\end{figure}

The property $\sigma_{\chi\overline{\chi}}\propto(s-4m^2_\chi)^{1/2}$ of the annihilation 
cross sections in the CP even fermionic Higgs portal model differs from
the corresponding property $\sigma_{DD}\propto(s-4m^2_D)^{-1/2}$, $D\in\{S,V\}$, of 
the bosonic models. This arises as a consequence of averaging over initial 
spins in the CP even model (\ref{eq:Lchi})
(with $k\equiv|\bm{k}|$),
\begin{equation}\label{eq:evensum}
\frac{1}{4}\sum_{s_1,s_2}
\left|\overline{v}(-\bm{k},s_2)\cdot u(\bm{k},s_1)\right|^2
=2k^2=\frac{1}{2}(s-4m^2_\chi).
\end{equation}
This leads to thermally averaged cross sections $\langle\sigma v\rangle(T)$ for 
given coupling $v_h/M$ in the CP even model (\ref{eq:Lchi}) which are considerably smaller
than the corresponding cross sections for given couplings $g_D$ in the bosonic models.
Comparison with the required cross section $\langle\sigma v\rangle_f$ for 
thermal freeze-out therefore leads to a considerably larger 
value for $v_h/M\gg g_D$ for given dark matter mass $m_\chi=m_D$, and this leads to
considerably larger recoil cross sections,
\begin{equation} \label{eq:recoilchiN}
\sigma_{\chi N}=\frac{g_{hN}^2v^2_h}{\pi M^2m_h^4}
\left(\frac{m_\chi m_N}{m_\chi+m_N}\right)^2,
\end{equation}
which are ruled out due to the limits from PandaX-II, LUX and XENON1T even for the 
weakest conceivable Higgs-nucleon coupling $g_{hN}v_h=210$ MeV, 
see e.g. Fig.~\ref{fig:chiheavy}.

However, the CP even model can be combined with the CP odd model in such a
way that the recoil cross sections still comply with the direct search limits.
This is discussed after the CP odd coupling in Eq. (\ref{eq:zeta}) below.

\subsection{CP odd coupling}

 Fermionic dark matter with a CP odd Higgs portal,
\begin{eqnarray}\nonumber
\mathcal{L}&=&\left[
\overline{\chi}\left(\mathrm{i}\gamma^\mu\partial_\mu-m_\chi\right)\chi
-\frac{\mathrm{i}}{\mu}\overline{\chi}\gamma_5\chi 
\left(H^+\cdot H-\frac{v_h^2}{2}\right)\right]_{h}
\\ \label{eq:Lchi5}
&=&\overline{\chi}\left(\mathrm{i}\gamma^\mu\partial_\mu-m_\chi\right)\chi
-\mathrm{i}\frac{v_h}{\mu} \overline{\chi}\gamma_5\chi h
-\frac{\mathrm{i}}{2\mu}\overline{\chi}\gamma_5\chi  h^2,
\end{eqnarray}
yields annihilation cross sections
\begin{equation}\label{eq:55tohh}
\sigma_{\chi\overline{\chi}\to hh}(s)=
\frac{\sqrt{s-4m_h^2}}{64\pi\mu^2\sqrt{s-4m_\chi^2}}
\frac{(s+2m_h^2)^2}{(s-m_h^2)^2},
\end{equation}
\begin{equation}\label{eq:55toff}
\sigma_{\chi\overline{\chi}\to f\overline{f}}(s)
=N_c\frac{\sqrt{s-4m_f^2}^3}{
16\pi\mu^2\sqrt{s-4m_\chi^2}}\frac{m_f^2}{
(s-m_h^2)^2+m_h^2\Gamma_h^2},
\end{equation}
\begin{equation}\label{eq:55toVV}
\sigma_{\chi\overline{\chi}\to ZZ,W^+ W^-}(s)
=\frac{\sqrt{s-4m_{W,Z}^2}}{32\pi\mu^2(1+\delta_z)\sqrt{s-4m_\chi^2}}
\frac{(s-2m_{W,Z}^2)^2+8m_{W,Z}^4}{
(s-m_h^2)^2+m_h^2\Gamma_h^2}.
\end{equation}

\begin{figure}[hptb]
\centerline{\psfig{file=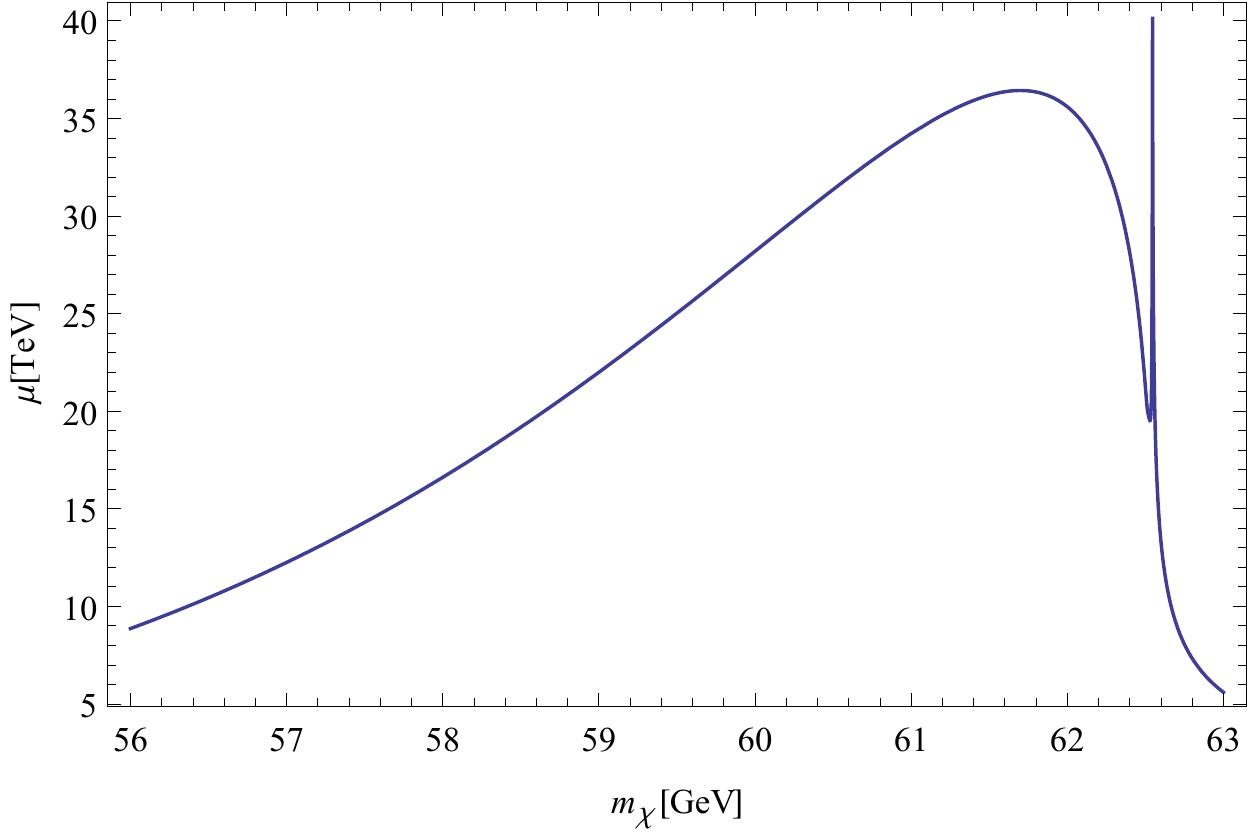,width=8cm}}
\vspace*{8pt}
\caption{The required coupling scale $\mu$ for dark matter creation through
the fermionic Higgs portal (\ref{eq:Lchi5}) in the WIMP mass range $53\,\mathrm{GeV}
\le m_\chi\le 63\,\mathrm{GeV}$. \label{fig:mu56to63} }
\end{figure}

The coupling scale $\mu$ can be determined from the requirement that thermal freeze-out
of the fermionic Higgs portal matter creates the observed dark matter abundance.
It varies between about 1 TeV and 40 TeV in the mass 
range $56\,\mathrm{GeV}\le m_\chi\le 63\,\mathrm{GeV}$, see Fig.~\ref{fig:mu56to63}

\begin{figure}[hptb]
\centerline{\psfig{file=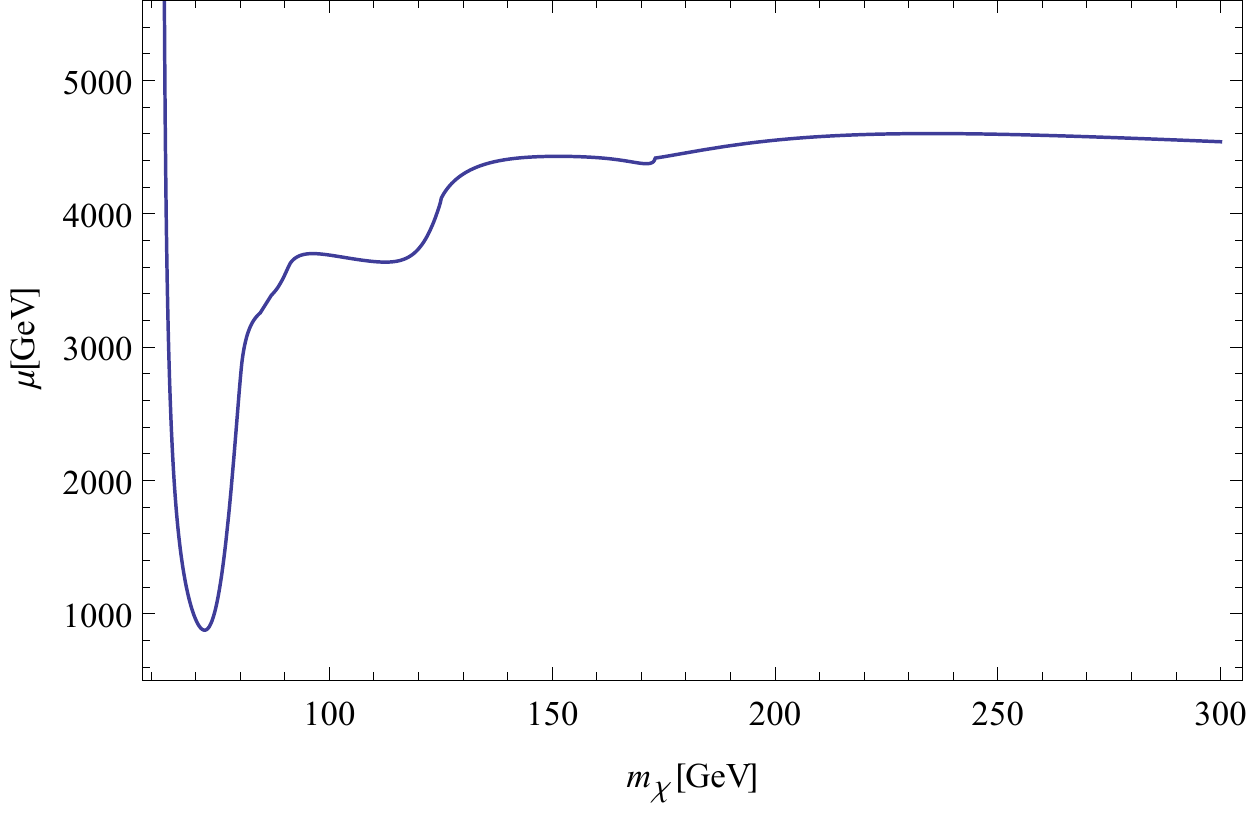,width=8cm}}
\vspace*{8pt}
\caption{The required coupling scale $\mu$ for dark matter creation through
the fermionic Higgs portal (\ref{eq:Lchi5}) in the WIMP mass range $63\,\mathrm{GeV}
\le m_\chi\le 300\,\mathrm{GeV}$. \label{fig:mu63to300} }
\end{figure}

The required scale $\mu$ increases near mass thresholds for $\chi\overline{\chi}$
annihilation, because opening up new annihilation channels implies that thermal freeze-out
can produce the observed dark matter abundance with smaller 
coupling $g_{h\chi}=v_h/\mu$, see Fig.~\ref{fig:mu63to300}. 
Above the top mass, $\mu$ decreases logarithmically with increasing dark matter 
mass $m_\chi$ for the same reason why the scale $M$ increased in the CP even model.

The required coupling scale for the CP odd model is always larger than the
corresponding scale for the CP even model, $\mu>M$. This is a consequence of the
pole near $s=4m^2_\chi$ in the annihilation cross sections (\ref{eq:55tohh}-\ref{eq:55toVV})
of the CP odd model. The CP odd dark fermion model behaves more like the bosonic models
in terms of the pole structure of the annihilation cross sections.
The reason for this different behavior of the CP even and odd fermionic models is 
that averaging over initial spins yields (\ref{eq:evensum}) 
in the CP even model (\ref{eq:Lchi}), whereas for the CP odd coupling 
in Eq. (\ref{eq:Lchi5}) we find
\begin{equation}\label{eq:oddsum}
\frac{1}{4}\sum_{s_1,s_2}
\left|\overline{v}(-\bm{k},s_2)\cdot\gamma_5\cdot u(\bm{k},s_1)\right|^2
=2\omega^2(\bm{k})=s/2.
\end{equation}
This also yields the narrow resonance at $m_\chi=m_h/2$
in Fig.~\ref{fig:mu56to63} for the CP odd coupling, which does not
appear for the CP even coupling in Fig.~\ref{fig:m56to63}.
The value of $\mu$ near the unitarity limit $m_\chi\simeq 100$ TeV
is $\mu\simeq 3.9$ TeV.

The nuclear recoil cross section differs from the corresponding 
result (\ref{eq:recoilchiN}) for the parity conserving coupling by 
a factor $\beta_\chi^2/2$, where $0\le\beta_\chi<1$ is the speed of the 
dark fermions,
\begin{equation}\label{eq:recoil5N}
\sigma_{\chi N}=\frac{g_{hN}^2v_h^2 m_N^2}{2\pi\mu^2 m_h^4}\frac{k^2}{
(m_N+m_\chi)^2}=\frac{g_{hN}^2v_h^2 m_N^2}{2\pi\mu^2 m_h^4}\frac{\beta_\chi^2 m^2_\chi}{
(m_N+m_\chi)^2}
\end{equation}
Unfortunately, this model is not testable through the direct search experiments
in the allowed mass 
range $56\,\mathrm{GeV}\lesssim m_\chi\lesssim 100\,\mathrm{TeV}$ because 
$\beta_\chi\sim 10^{-3}$ implies that the nuclear recoil cross sections for this
model are below the neutrino floor\cite{billard} for all possible Higgs-nucleon
couplings (\ref{eq:gvh5}).

This difference between the CP even and odd couplings is again due to the very
different behavior under averaging over initial spins and summing over final
spins for the recoil events. For the CP even model, the result
\begin{equation}\label{eq:spinaverage2}
\frac{1}{2}\sum_{s,s'}\left|\overline{u}(\bm{p},s')\cdot u(\bm{k},s)\right|^2
=\frac{1}{2}\mathrm{tr}\!\left[(m-\gamma\cdot p)(m-\gamma\cdot k)\right]
=2\left(m^2-p\cdot k\right)
\end{equation}
yields a factor $4m^2$ in the non-relativistic limit, whereas for the CP odd
coupling the result
\begin{eqnarray}\nonumber
\frac{1}{2}\sum_{s,s'}\left|\overline{u}(\bm{p},s')\cdot\gamma_5\cdot u(\bm{k},s)\right|^2
&=&-\frac{1}{2}\mathrm{tr}\!\left[\gamma_5\cdot(m-\gamma\cdot p)
\cdot\gamma_5\cdot(m-\gamma\cdot k)\right]
\\  \label{eq:spinaverage3}
&=&-\,2\left(m^2+p\cdot k\right)
\end{eqnarray}
yields a factor $(\bm{p}-\bm{k})^2$ in the non-relativistic limit.

\begin{figure}[hptb]
\centerline{\psfig{file=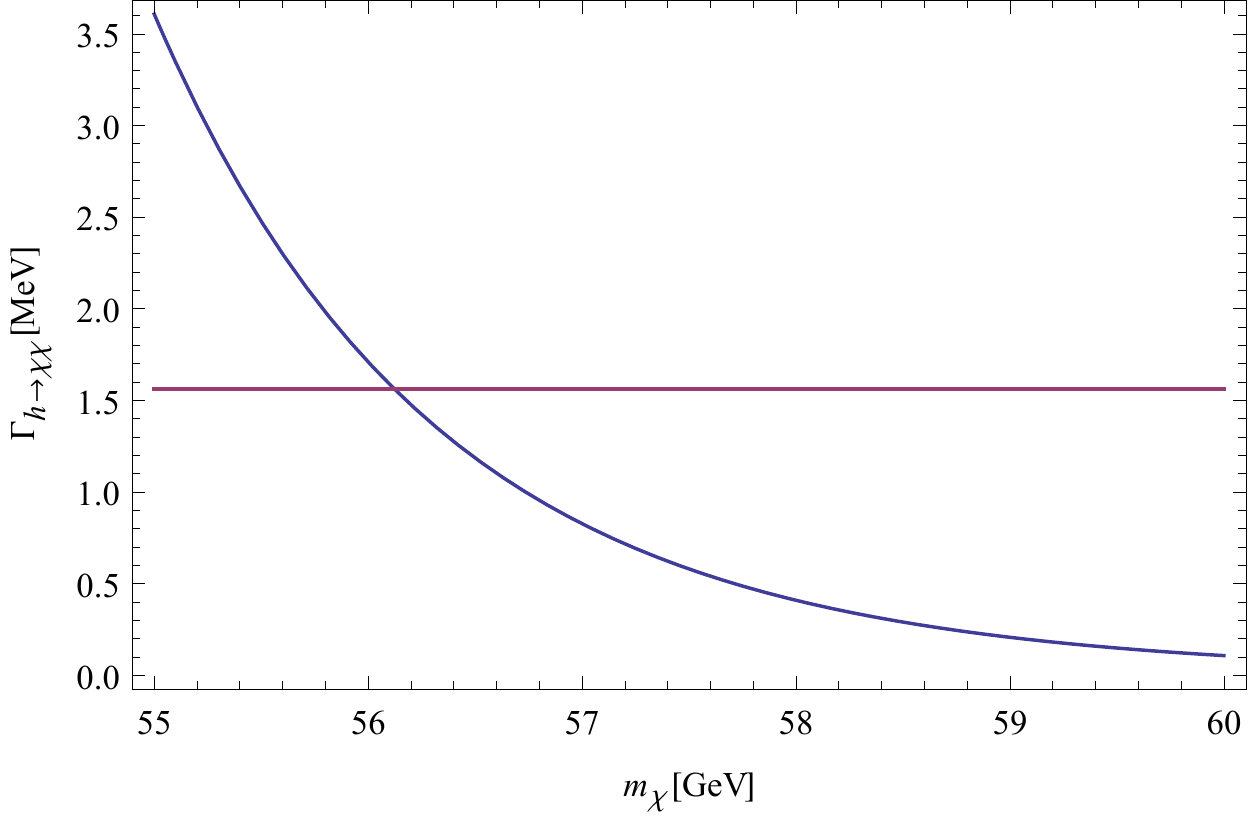,width=8cm}}
\vspace*{8pt}
\caption{The invisible decay width $\Gamma_{h\to\chi\overline{\chi}}$ for 
the CP odd coupling (\ref{eq:Lchi5}) and $m_\chi$
between 55 GeV and 60 GeV. The horizontal line arises from the limit on the 
branching ratio for invisible Higgs 
decays, $\mathcal{B}=\Gamma_{h\to\mathrm{inv.}}/(\Gamma_{h\to\mathrm{inv.}}
+\Gamma_{h\to\mathrm{SM}})\le 0.24$. The mass parameter $M$ for the $h\chi^2$ coupling 
is determined from the requirement that the fermionic Higgs portal matter accounts for 
the observed dark matter. \label{fig:hto55} }
\end{figure}

The small recoil cross sections imply that the constraint from the Higgs decay 
width for $m_\chi<m_h/2$,
\begin{equation}\label{eq:hto55}
\Gamma_{h\to\chi\overline{\chi}}
=\frac{v_h^2}{8\pi\mu^2}\sqrt{m_h^2-4m_\chi^2},
\end{equation}
is significant for this dark matter model and
implies $m_\chi\gtrsim 56.1$ GeV, see Fig.~\ref{fig:hto55}.

The partial decay widths and the resulting mass constraints are virtually identical for 
the CP even coupling (\ref{eq:Lchi}) and the CP odd coupling (\ref{eq:Lchi5}) because 
the relative factor $s/(s-4m_\chi^2)$ between the respective annihilation cross 
sections (\ref{eq:chichi2hh}-\ref{eq:chichi2VV}) and (\ref{eq:55tohh}-\ref{eq:55toVV}) 
near $m_\chi\lesssim m_h/2$ enters the calculation of the coupling scales $M$ 
versus $\mu$ and compensates for the relative factor
$m_h^2/(m_h^2-4m^2_\chi)$ between the partial decay widths (\ref{eq:h2chichi})
and (\ref{eq:hto55}). For example, the partial decay width (\ref{eq:h2chichi}) for 
the CP even model at $m_\chi=55$ GeV is $\Gamma_{h\to\chi\overline{\chi}}=3.76$ MeV,
while the corresponding contribution to the Higgs decay width for the CP odd
model is $\Gamma_{h\to\chi\overline{\chi}}=3.61$ MeV.

The CP even and odd couplings can be combined,
\begin{equation}\label{eq:zeta}
\mathcal{H}_{\chi h}=
\overline{\chi}\cdot\!\left(\frac{1}{M_\zeta}+\frac{\mathrm{i}}{\mu_\zeta}\gamma_5\right)
\!\cdot\chi\left(H^+H-\frac{v_h^2}{2}\right),
\end{equation}
with $M_\zeta=M/\zeta$, $\mu_\zeta=\mu/\sqrt{1-\zeta^2}$, such that the model still
complies with the direct search limits. 

The CP even and CP odd amplitudes do not interfere since spin averaging in the 
interference terms leaves terms proportional 
to tr$(\gamma_5\gamma_\mu\gamma_\nu)=0$, tr$(\gamma_5\gamma_\mu)=0$, 
and tr$(\gamma_5)=0$.
The annihilation cross sections and nulear recoil cross sections
are therefore the sums of the cross sections
of the CP even $(+)$ and CP odd $(-)$ 
models, $\sigma=\zeta^2\sigma_{(+)}+(1-\zeta^2)\sigma_{(-)}$,
and the same applies to the invisible Higgs decay width,
$\Gamma_{h\to\chi\overline{\chi}}=\zeta^2\Gamma^{(+)}_{h\to\chi\overline{\chi}}
+(1-\zeta^2)\Gamma^{(-)}_{h\to\chi\overline{\chi}}$.

\begin{figure}[hptb]
\centerline{\psfig{file=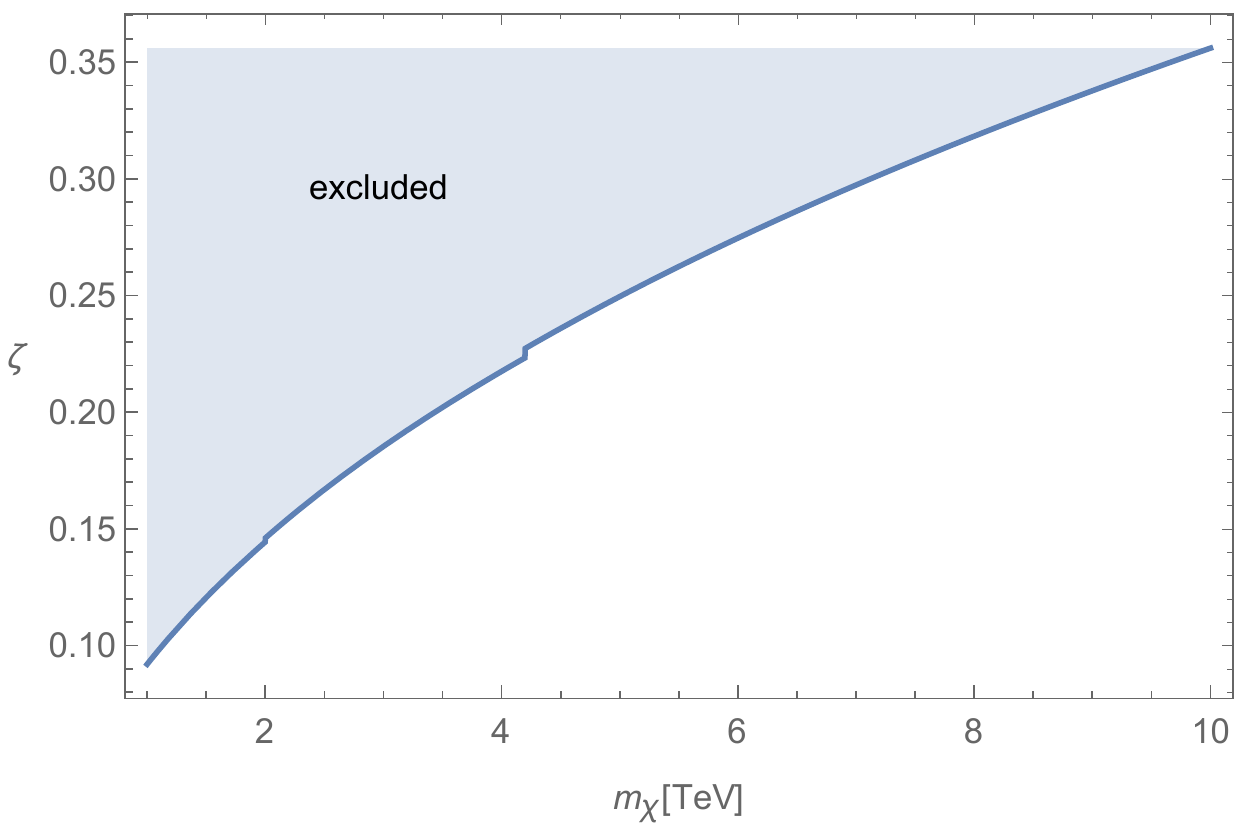,width=8cm}}
\vspace*{8pt}
\caption{The maximal values of $\zeta$ for which the Higgs portal
coupling (\ref{eq:zeta}) still complies with extrapolated XENON1T 
limits in the mass range between 1 TeV and 10 TeV.\label{fig:zeta} }
\end{figure}

The property for the annihilation cross sections also implies that the required 
values of the coupling parameters $M(m_\chi)$ and $\mu(m_\chi)$ 
do not depend on $\zeta$: The freeze-out requirements for $\zeta=1$,
 \textit{viz.} $\langle\sigma v\rangle_f=\langle\sigma v\rangle_{(+)}(M,T_f)$,
and for $\zeta=0$, 
 \textit{viz.} $\langle\sigma v\rangle_f=\langle\sigma v\rangle_{(-)}(\mu,T_f)$,
imply 
\begin{equation}\label{eq:zetafreeze}
\langle\sigma v\rangle_f=\zeta^2\langle\sigma v\rangle_{(+)}(M,T_f)
+(1-\zeta^2)\langle\sigma v\rangle_{(-)}(\mu,T_f)
\end{equation}
 for every value of $\zeta$. $\zeta$ can therefore be directly determined from 
the requirement that $\sigma_{\chi N}$ complies with the limits from the direct
search experiments. Maximal values of $\zeta$ which comply with the XENON1T 
constraints and for Higgs-nucleon coupling $g_{hN}v_h=210$ MeV are displayed 
in Fig.~\ref{fig:zeta}. Increasing the Higgs-nucleon coupling to $g_{hN}v_h=289$ 
MeV reduces the allowed values of $\zeta$ by a factor 0.727.

 For better or worse, this defines a one-parameter Higgs portal model which
can always be arbitrary close to the direct search limits, and yet also always 
be safe from being ruled out by direct search experiments.

\section{Conclusions}	
\label{sec:conc}

Relic electroweak singlets with a
Higgs portal coupling provide an interesting scenario for
dark matter due to the direct connection between mass and coupling to baryons,
which implies high predictivity and therefore also generically high testability
of the models. This in turn also implies easy verification of any potential 
direct signal through indirect searches. 

In particular, the fermionic singlet Higgs portal model with mass suppressed 
purely CP even coupling to the Higgs field appears to be ruled out now due 
to the limits from
the direct search experiments. Bosonic models are constrained to dark matter masses
near the resonance region, $57\,\mathrm{GeV}\lesssim m_D\lesssim m_h/2$, or to
heavy dark matter masses in the few TeV range or higher. They can be tested up to the
unitarity limit by DARWIN and likely also by DarkSide-20k, while XENONnT and LZ
will already have the potential to test the bosonic models up to several ten TeV.
The fermionic singlet model with pure mass supressed CP odd Higgs coupling
is constrained by the limits on the invisible Higgs decay width to masses
$m_\chi\gtrsim 56$ GeV, but the nuclear recoil cross sections proportional to
$\beta_\chi^2$ are below the neutrino floor and therefore evade the direct search
experiments. 

The bosonic electroweak singlet Higgs portal models are increasingly constrained
by current direct search experiments, and will be further tested by direct
search experiments which are currently under construction. Their highly
predictive features and their close alignment with current and future sensitivities
of direct search experiments make them interesting targets for experimental dark 
matter research, since any Higgs portal interpretation of a direct detection signal 
can easily be tested with follow-up cosmic ray observations for the bosonic models.
The fermionic Higgs portal with CP odd coupling is an interesting target for collider
based searches and indirect searches, but relevance for the direct search experiments
requires admixture of a CP even Higgs coupling.

\section*{Acknowledgments}

This work was supported by the Natural Sciences and Engineering
Research Council of Canada. I would like to thank Jose Alarc\'on Soriano
and Jusak Tandean for very helpful and constructive comments.

\end{document}